\documentclass[%
 reprint, %10.5pt,%onecolumn,
 amsmath,amssymb,
 aps,
]{revtex4-2}

\usepackage{graphicx}% Include figure files
\usepackage{dcolumn}% Align table columns on decimal point
\usepackage{bm}% bold math
\usepackage{float}% 
\usepackage{here}% 

\usepackage[dvipsnames]{xcolor}
\usepackage{stackengine}
\def\subrangle#1{\stackengine{11.5pt}{}{$\!\scriptstyle #1$}{U}{l}{F}{F}{L}}

\newcommand{\vect}[1]{\boldsymbol{\mathbf{#1}}}
\newcommand{\ER}{Erd\H{o}s-R\'{e}nyi }
\newcommand{\vecm}{{\bf m}}

\newcommand{\papertitle}{Financial fire sales as continuous-state complex contagion}

\begin{document}

%\title{multistate complex contagion}% Force line breaks with \\
%\title{multistate complex contagion in financial systems}% 
%\title{Financial fire sales as multistate complex contagion}% 
\title{\papertitle}% 
%\title{multistate complex contagion in financial markets}% 

\author{Tomokatsu Onaga}
%\email{onaga@tohoku.ac.jp}
\affiliation{
Frontier Research Institute for Interdisciplinary Sciences, Graduate School of Information Sciences,\\ Tohoku University, Sendai, Japan
}

\author{Fabio Caccioli}%
% \email{}
\affiliation{%
  Department of Computer Science, University College London, London, UK\\ Systemic Risk Centre, London School of Economics and Political Science, London, UK\\
London Mathematical Laboratory, London, UK
}%
\author{Teruyoshi Kobayashi}%
 \email{kobayashi@econ.kobe-u.ac.jp}
\affiliation{%
  Department of Economics, \\ Center for Computational Social Science, Kobe University, Kobe, Japan
}%

\date{\today}

\begin{abstract}
Trading activities in financial systems create various channels through which systemic risk can propagate. An important contagion channel is financial fire sales, where a bank failure causes asset prices to fall due to asset liquidation, which in turn drives further bank defaults, triggering the next rounds of liquidation. 
This process can be considered as complex contagion, yet it cannot be modeled using the conventional binary-state contagion models because there is a continuum of states representing asset prices. 
Here, we develop a threshold model of continuous-state cascades in which the states of each node are represented by real values. We show that the solution of a multi-state contagion model, for which the continuous states are discretized, accurately replicates the simulated continuous state distribution as long as the number of states is moderately large. 
This discretization approach allows us to exploit the power of approximate master equations (AME) to trace the trajectory of the fraction of defaulted banks and obtain the distribution of asset prices that characterize the dynamics of fire sales on asset-bank bipartite networks.
We examine the accuracy of the proposed method using real data on asset-holding relationships in exchange-traded funds (ETFs).
\end{abstract}

\maketitle

\section{introduction}
Financial markets are complex systems in which many economic agents, such as individuals, firms, financial institutions, and central banks, are interconnected at the global scale through trading activities. 
The networks of money and credit flows have been expanding as more market participants have access to a greater variety of financial assets, bonds, and other securities. 
On one hand, this increases the efficiency of the market. On the other hand, the past two decades made it clear that increased connectivity between market participants leads to higher risk of systemic failures in the global financial market: Once a part of the financial system malfunctions, interconnections can facilitate a chain reaction that can spread globally  \cite{Arinaminpathy2012PNAS,Battiston2016PNAS,Battiston2016Science}, for instance in the form of fire sales of financial assets~\cite{Cifuentes2005JEEA,caccioli2014JBF,Caccioli2015}, or of cascades of defaults among banks with mutual exposures~\cite{GaiKapadia2010,Cont2013,amini2016resilience,Brummitt2015PRE,hurd2016Book}.

In models of financial contagion, nodes can be financial institutions, individual traders, or financial assets, depending on which aspects of financial phenomena one would examine.
For example, in cascades of bank defaults due to counterparty default risk, nodes are banks and edges are lending and borrowing relationships in an interbank market~\cite{GaiKapadia2010,Cont2013,fouque2013handbook,kobayashi2013network_vs,Brummitt2015PRE,hurd2016Book,caccioli2018review}.
In cascades of financial fire sales, nodes are financial assets and financial institutions, which form a bipartite network defined by asset-holding relationships, i.e., overlapping portfolios~\cite{Huang2013SciRep,caccioli2014JBF,Caccioli2015,vodenska2021systemic}. 

However, full analytical solutions for these models are difficult to obtain since the states of nodes in financial networks are generally non-binary unlike the standard threshold models~\cite{Watts2002,Gleeson2007,Gleeson2008,GaiKapadia2010,Brummitt2015PRE,kobayashi2015trend,burkholz2016damage,unicomb2019reentrant,gleeson2018message,kobayashi2021dynamics}.
% Both types of financial cascades have been extensively studied within the class of binary-state threshold models, which were originally developed for the study of information cascades in social networks~\cite{Watts2002,Watts2007}.
%The states of nodes in financial networks, however, are not inherently binary. 
While the state of a bank can for instance be well captured by a binary variable that denotes whether the bank is solvent or insolvent, the state of an asset (e.g., a stock price) is a continuous variable, since in principle its price can take any non-negative real value.
Financial fire sales are thus continuous-state complex contagion processes in which a group of nodes are characterized by real-valued states, while the others by binary states.

%To generalize the conventional binary-state cascade models, we develop a threshold model of continuous-state contagion in which the states of each node are represented by real values. The difficulty, however, is that the distribution of nodes' states is generally given as a density function, an infinite object, as opposed to a discrete histogram. We therefore employ a discretization approach in which a set of real-valued states is approximated by a finite number of discretized states. As a tool for the calculation of diffusion dynamics, a \emph{message-passing method} has been used in the analysis of binary-state cascades~\cite{Gleeson2007,Gleeson2008,gleeson2018message}. 
Here we develop a threshold model of continuous-state contagion in which the states of each node are represented by real values.
Compared to the conventional binary-state cascade models, the difficulty is that the distribution of nodes' states is generally given as a 
density function as opposed to a probability mass function. 
We therefore employ a discretization approach in which we approximate a set of real-valued states  by a finite number of discretized states. 

As a tool for the calculation of spreading dynamics, a \emph{message-passing method} has been used in the analysis of binary-state cascades~\cite{Gleeson2007,Gleeson2008,gleeson2018message,kobayashi2021dynamics}. 
While the message-passing method is highly accurate in describing binary-state spreading processes~\cite{gleeson2018message}, it can be imprecise in models with more than two states, i.e., \emph{multistate} cascade models~\cite{Melnik2013,fennell2019multistate}.
We address this problem by employing a more general \emph{approximate master equations} (AMEs) approach~\cite{gleeson2011high,gleeson2013binary,fennell2019multistate}, in which the states of neighboring nodes are not necessarily independent unlike the conventional mean-field approaches. We show that the AME solution well explains the distribution of continuous states in the generalized threshold model as long as the number of discretized states is moderately large. 

By extending the continuous-state threshold model, we develop an analytical framework to study the dynamics of financial fire sales due to overlapping portfolios. In the model, \emph{bank nodes} take binary states (i.e., solvent or insolvent) while the states of \emph{asset nodes} (i.e., asset prices) are represented by real values.
The analysis reveals that the discrete-price (i.e., discrete-state) equilibrium obtained by the AME method converges to the continuous-price (i.e., continuous-state) equilibrium as one increases the number of discretized states.  This suggests that the behavior of continuous asset prices in the propagation of fire sales, which has been examined only numerically in previous studies~\cite{Huang2013SciRep,Caccioli2015,vodenska2021systemic}, can be analyzed analytically within a framework of multistate cascade models.

\section{Continuous-state complex contagion}\label{sec:continuous_watts}

\subsection{A threshold model with continuous states}

We first present a model of continuous-state contagion in which the state of each node is represented by a real value $s\in[0,1]$. 
We will then consider a model of financial fire sales with both binary and continuous states in section III.

The model we consider in this section is a generalized version of the binary-state Watts' threshold model~\cite{Watts2002}; in addition to state~0 (i.e., fully inactive) and state~1 (i.e., fully active), there is a spectrum of intermediate states between them (i.e., partially active). 

%Let $\vect{s}=(s_1,\ldots,s_k)^\top$ be the vector of neighbors' states where its length equals the degree $k$ of the focal node.
Let us focus on a node with degree $k$. We will denote by $\vect{s}=(s_1,\ldots,s_k)^\top$ the vector containing the states of the node's neighbors.
While there can be various specifications of continuous-state cascades, we consider a generalized fractional threshold rule in which  the state of each node is affected by the fraction of its ``active neighbors'' $m/k$, as it is assumed in many variants of the discrete-state threshold model~\cite{Watts2002,ikeda2010cactus,Melnik2013,Nematzadeh2014,kobayashi2015trend,Brummitt2015PRE,Karimi2013PhysicaA,ruan2015kinetics,gleeson2018message,unicomb2019reentrant,unicomb2021dynamics}. For a given threshold $\theta\in[0,1)$, the response function $\mathcal{F}_{\vect{s}}$ is generically given as:
\begin{align}
    \mathcal{F}_{\vect{s}} =  
    \begin{cases}
   f\left( \frac{m}{k}\right)\; {\text{ if } }\;  \frac{m}{k} > \theta , \\
         0 \;\text{ otherwise},
     \end{cases}
    \label{eq:F_continuous_watts}
\end{align}
where $m$ denotes the number of ``active'' neighbors whose states are positive: $m=|\{j:s_j>0,s_j\in\vect{s}\}|$. $f$ is a continuous non-decreasing function on $(\theta,1]$ such that $f(\cdot)\in[0,1]$. That is, $\mathcal{F}_{\vect{s}}$ gives the state of a node, taking the neighbors' states $\vect{s}$ as input. 
Note that the standard Watts model is recovered when $f(m/k) = 1$ for $m/k>\theta$, corresponding to the most ``progressive'' response function. 
In Appendix~\ref{sec:threshold_discretization}, we examine a more general nonlinear form of $f(m/k)$.

We also consider an alternative response function that depends on the sum of neighbors' states:
\begin{align}
      \widetilde{\mathcal{F}}_{\vect{s}} =  
    \begin{cases}
   f\left( \frac{\sum_j s_j}{k}\right)\; {\text{ if } }\;  \frac{\sum_j s_j}{k} > \theta ,  \\
         0 \;\text{ otherwise}.
     \end{cases}
    \label{eq:F_continuous_watts_sum}
\end{align}
It should be noted that the influence of seed nodes will generally be weakened when the response function is given by $\widetilde{\mathcal{F}}_{\vect{s}}$ rather than $\mathcal{F}_{\vect{s}}$. 
For example, consider a tree in which each node (excluding the seed node) has $k$ edges, and all nodes are initially in state $0$.
Suppose that $f(\sum_j s_j/k)=\sum_j s_j/k$, and the state of the seed node shifts from $0$ to $1$. Since the seed node is one of the $k$ neighbors for the child nodes, the states of the child nodes become $1/k$ if $\theta < 1/k$. Then, the states of the grandchild nodes become $1/k^2$ if $\theta < 1/k^2$, the great-grandchild nodes' state become $1/k^3$ if $\theta < 1/k^3$, and so on. While the actual propagation process can be more complicated due to repercussion of peer effects, the spread of influence with response function $\widetilde{\mathcal{F}}_{\vect{s}}$ is more likely to decay than that obtained with the response function $\mathcal{F}_{\vect{s}}$.

\subsection{Multistate dynamical process as an approximation }
\label{sec:multistate AME}
 In the continuous-state contagion model, it is difficult to describe the collective dynamics in an analytical manner because the shares of nodes in each state are represented as a continuous density function in the limit of an infinitely large network size. 
 While the conventional mean-field/message-passing methods could be applied to a class of multi-stage contagion models with several nodal states, the accuracy of the approximation methods easily deteriorates when there are more than two states~\cite{Melnik2013}.
 
To address this problem, we introduce a discretization approach using an approximate master equation (AME) method~\cite{gleeson2011high,gleeson2013binary,fennell2019multistate}. To apply the AME method, we transform the continuous-state model into a discrete multistate model by discretizing the range of continuous states into $n$ grids. 
Let $\mathbf{m}\equiv (m_0, m_1, \ldots, m_{n-1})^\top$ be the profile of neighbors' states, where $m_\ell$ denotes the number of neighbors in (discretized) state $\ell\in\{0,\ldots,n-1\}$. For a node with degree $k\in\{1,\ldots ,k_{\rm max}\}$, we have $\sum_{\ell=0}^{n-1}m_\ell=k$. 

Transitions in the states of nodes are captured by the response function $F_\mathbf{m}(i\rightarrow j)$, the rate at which a node in state $i$ changes its state to $j$.
For a given sequence of thresholds $\Theta=\{\theta_0,\theta_1,\ldots,\theta_{n-1}\}$,  $F_\mathbf{m}$ is specified as
\begin{equation}
    F_\mathbf{m}(i\rightarrow j) =
    \begin{cases}
    1 \;\text{ if } \theta_{j-1}< P(\mathbf{m})\le \theta_j
    \;\; \\
    0 \;\text{ otherwise}
    \end{cases}
    \label{eq:F_multistate}
\end{equation}
for $1\le j \le n-1$, and
\begin{equation}
    F_\mathbf{m}(i\rightarrow 0) =
    \begin{cases}
    1 \;\text{ if }  P(\mathbf{m})\le \theta_0
    \;\; \\
    0 \;\text{ otherwise},
    \end{cases}
    \label{eq:F_multistate_j0}
\end{equation}
where $P(\mathbf{m})$ denotes the ``peer pressure'', i.e., the influence from neighbors, and $\theta_{j-1}$ and $\theta_j$ are the thresholds between which the node is in state $j$. 

To approximate the continuous-state model, the peer-pressure corresponding to the response function $\mathcal{F}_{\vect{s}}$ (Eq.~\ref{eq:F_continuous_watts}) is given as
\begin{equation}
    P(\mathbf m) = f\left(\frac{\sum_{\ell=1}^{n-1} m_\ell}{k}\right).
    \label{eq:response_function_P}
\end{equation}
Note that the multistate model reduces to a binary-state model when $n=2$, $0<\theta_0<1$ and $\theta_1=1$.
We specify the sequence of threshold values $\Theta$ such that the responsiveness $f(m/k)$ in the continuous model is evenly spaced on $[0,1]$, where $f(\theta_j)-f(\theta_{j-1}) = 1/(n-1)$ for all $j=1,\ldots,n-1$. 
A detailed description of the specifications of threshold values is provided in Appendix~\ref{sec:threshold_discretization} and Fig.~\ref{fig:schematic_response_SI}.
When the response function is given by $\widetilde{\mathcal{F}}_{\vect{s}}$ (Eq.~\ref{eq:F_continuous_watts_sum}), the peer pressure for the discretized model leads to
\begin{align}
    P(\mathbf m) = f\left(\frac{1}{k}\frac{\sum_{\ell=1}^{n-1}\ell m_\ell}{n-1}\right),
    \label{eq:response_function_P_sum}
\end{align}
where $\ell/(n-1)\in [0,1]$ is the normalized state of neighbors in state $\ell$. 
In the following, we specify the functional form of $f$ as $f(u)=\left(\frac{u-\theta}{1-\theta}\right)^\eta$ with $\eta\geq 0$, where $\eta=0$ corresponds to a linear threshold model (see Appendix~\ref{sec:threshold_discretization} for details).

Let $x_{k,{\bf m}}^i(t)$ be the fraction of $k$-degree nodes in state $i$ that face neighbors' profile $\bf m$ at time $t$.
Employing the AME formalism, we solve the following differential equations for the dynamics of $x_{k,{\bf m}}^i(t)$~\cite{fennell2019multistate}:
\begin{eqnarray}
    \frac{dx^i_{k,\mathbf{m}}}{dt} &=& - \sum_{j>i}F_\mathbf{m}(i\rightarrow j) x^i_{k,\mathbf{m}}
    + \sum_{j<i}F_\mathbf{m}(j\rightarrow i) x^j_{k,\mathbf{m}} \nonumber\\
    && - \sum_{l=0}^{n-1}\sum_{\ell'>\ell} m_\ell\beta^i(\ell\rightarrow \ell') x^i_{k,\mathbf{m}} \nonumber \\
    && + \sum_{l=0}^{n-1}\sum_{\ell'<\ell} (m_{\ell'}+1)\beta^i(\ell'\rightarrow \ell) x^i_{k,\mathbf{m}-\mathbf{e}_\ell+\mathbf{e}_{\ell'}},
     \label{eq:AME_eq}
\end{eqnarray}
where ${\bf e}_\ell$ denotes the $n\times 1$ vector that contains $1$ for the $\ell$-th element and $0$ for the other elements. 
$\beta^{i}(\ell\to\ell')$ denotes the probability that a neighbor of a node being in state $i$ changes its state from $\ell$ to $\ell'$:
\begin{equation}
\beta^i(\ell\rightarrow \ell') = \frac{
\langle \sum_{|\mathbf{m}|=k} m_i x^\ell_{k,\mathbf{m}}(t) F_\mathbf{m}(\ell\rightarrow \ell') \rangle_k
}{
\langle \sum_{|\mathbf{m}|=k} m_i x^\ell_{k,\mathbf{m}}(t) \rangle_k
},\label{eq:AME_beta}
\end{equation}
where $\langle\cdot\rangle_k$ denotes the average over degree $k$ with degree distribution $p_k$.  
The first (resp. second) term of Eq.~\eqref{eq:AME_eq} captures the rate at which the states of a $k$-degree node in the $(i,\vecm)$ class (resp. $(j,\vecm)$ class) shifts from $i$ to $j (\neq i)$ (resp. $j$ to $i (\neq j)$) in an infinitesimal time interval $dt$. It should be noted that Eq.~\eqref{eq:AME_eq} describes the dynamics under \emph{asynchronous update} in which only a fraction $dt$ of nodes can change their states in response to their neighbor profiles in a small time interval~\cite{Gleeson2007, Melnik2013}.
The third term denotes the rate at which the neighbors' state profile will be different from $\vecm$.
The forth terms indicates the rate at which the neighbors' profile newly becomes $\vecm$. The expression $\vecm-{\bf e}_\ell+{\bf e}_{\ell'}$ represents the neighbor profile that has $m_{\ell'}+1$ in the $\ell'$-th element and $m_{\ell}-1$ in the $\ell$-th element.
For $k$ ranging from $0$ to $k_{\rm max}$, the total number of differential equations leads to $n\sum_{k=0}^{k_{\rm max}}\binom{k+n-1}{k}$~\cite{fennell2019multistate}.

\begin{figure*}[tb]
    \centering
    \includegraphics[width=17cm]{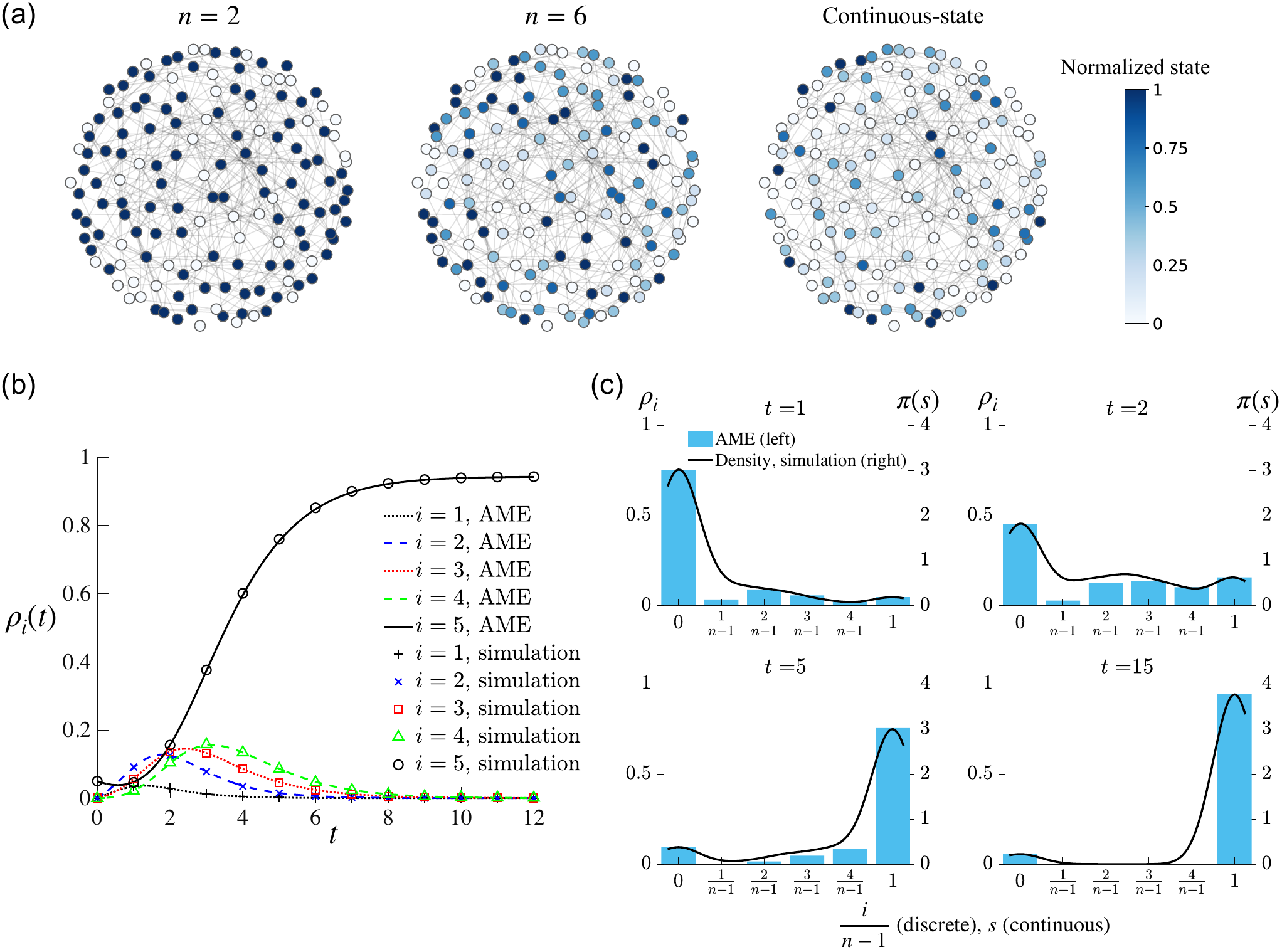}
    \caption{State transition in the continuous- and discrete-state threshold models. 
    (a) Visualization of node states at $t=3$ in different models. The network structure and the seed nodes are common for the three cases. 
    (b) Comparison between the AME solution and simulation results in the discrete-state model with $n=6$. 
    (c) Distribution of states at a given time. In the AME, the fraction of nodes in  discretized state~$i\in\{0,\ldots,n-1\}$ is given by $\rho_i$ (blue bar). For the continuous-state model, density $\pi(s)$ for state $s\in[0,1]$ is calculated by the kernel density estimation with bandwidth $0.1$ based on numerical simulations (black line). The index of discretized states is normalized by $n-1$ so that the states are distributed on $[0,1]$. Seed fraction is determined such that the fraction $0.05$ of nodes are in the ``most-active'' state (i.e., $i=n-1$ or $s=1$) while $0.95$ of nodes are in state~0. For panels b and c, we run 40 simulations on \ER networks with $N=3000$, $z=3$, $\theta=0$, and $\theta_j=j/(n-1)$. }
    \label{fig:continuous_watts}
\end{figure*}

\subsection{Validation}

To examine the accuracy of the discrete approximation, we calculate the average fraction of active nodes in a discretized state~$i$ at time $t$, denoted by $\rho_i(t)$, using the solutions of the AMEs~\eqref{eq:AME_eq}:
\begin{align}
    \rho_i(t) = \left\langle\sum_{|\mathbf{m}|=k} x^i_{k, \mathbf{m}}(t)\right\rangle\subrangle{k},
    \label{eq:rho_i_watts}
\end{align}
where $\sum_{|\mathbf{m}|=k}$ denotes the sum over all combinations of $\mathbf{m}$ such that $\sum_{\ell=0}^{n-1}m_\ell=k$. 
On the other hand, the states of nodes in the continuous-state model are represented by $\pi(s)$ on $s\in[0,1]$, where $\pi(s)$ is a density of nodes in state~$s$. We consider \ER random networks in which the degree distribution follows a Poisson distribution with mean $z$. 
We first consider a simple responsiveness given as $f(m/k)=m/k$ (i.e., $\eta=1$, $\theta=0$), which is a linear map of the fractional influence $m/k$. The corresponding thresholds are given by $\theta_j = j/(n-1)$ for $j=0,\ldots,n-1$. 
Results for more general nonlinear functional forms are presented in Appendix~\ref{sec:threshold_discretization}. 
For the calculation of the AME solution, we use the MATLAB code developed by~\cite{fennell2019multistate,FennelCode}.

While the states of nodes are $0$ or $1$ when $n=2$ (Fig.~\ref{fig:continuous_watts}a, \emph{left}), there are many nodes whose states are in between $0$ and $1$ when $n>2$ once the state index is normalized by $n-1$ (Fig.~\ref{fig:continuous_watts}a, \emph{middle}). 
In the discrete-state model with $n=6$, the node states are distributed as heterogeneously as in the continuous-state model (Fig.~\ref{fig:continuous_watts}a, \emph{middle} and \emph{right}). On the other hand, the steady-state fraction of nodes in state~1 in the binary-state approximation (i.e., $n=2$) is too large, compared to that of the continuous-state model (Fig.~\ref{fig:continuous_watts_theta018} in Supplemental Material (SM)).

As shown in previous studies, the AME method is highly accurate in predicting the dynamical path of the fractions of each state in a class of discrete-state contagion models~\cite{fennell2019multistate} (Figs.~\ref{fig:continuous_watts}b and \ref{fig:path_different_eta}). 
It is thus sufficient to compare $\rho_i$ calculated by the AMEs with the density function $\pi(s)$ obtained via numerical simulations of the continuous-state model. 
Indeed, the probability mass function for $\rho_i$ gives an accurate approximation to the continuous density function $\pi(s)$ (Fig.~\ref{fig:continuous_watts}c). In Figs.~\ref{fig:continuous_watts_theta018} and \ref{fig:continuous_watts_dist_eta} in SM, we show that the discrete approximation with $n=6$ still maintains its accuracy when $\theta>0$ and the responsiveness is more ``progressive'' (i.e., $f$ is concave, Fig.~\ref{fig:schematic_response_SI}a) or ``conservative'' (i.e., $f$ is convex, Fig.~\ref{fig:schematic_response_SI}c).
As expected, the probability mass function converges to the continuous density function $\pi$ as one increases the number of states $n$ (Fig.~\ref{fig:continuous_watts_theta018}). 

On the other hand, when the response function is given by $\widetilde{\mathcal{F}}_{\vect{s}}$, the AME method does not necessarily provide a good approximation. The AME method works well when $\eta<1$ and $f$ is concave, where the threshold values are skewed toward the origin as illustrated in Fig.~\ref{fig:schematic_response_SI}a, but otherwise it may incorrectly predict the shares of each state (Fig.~\ref{fig:continuous_AME_sum}).
This suggests that for the continuous model to be well approximated, the discretized thresholds need to be sufficiently densely distributed near $\theta (=\theta_0)$.

\section{A model of financial fire sales}

\begin{figure*}[tb]
    \centering
    \includegraphics[width=12.5cm]{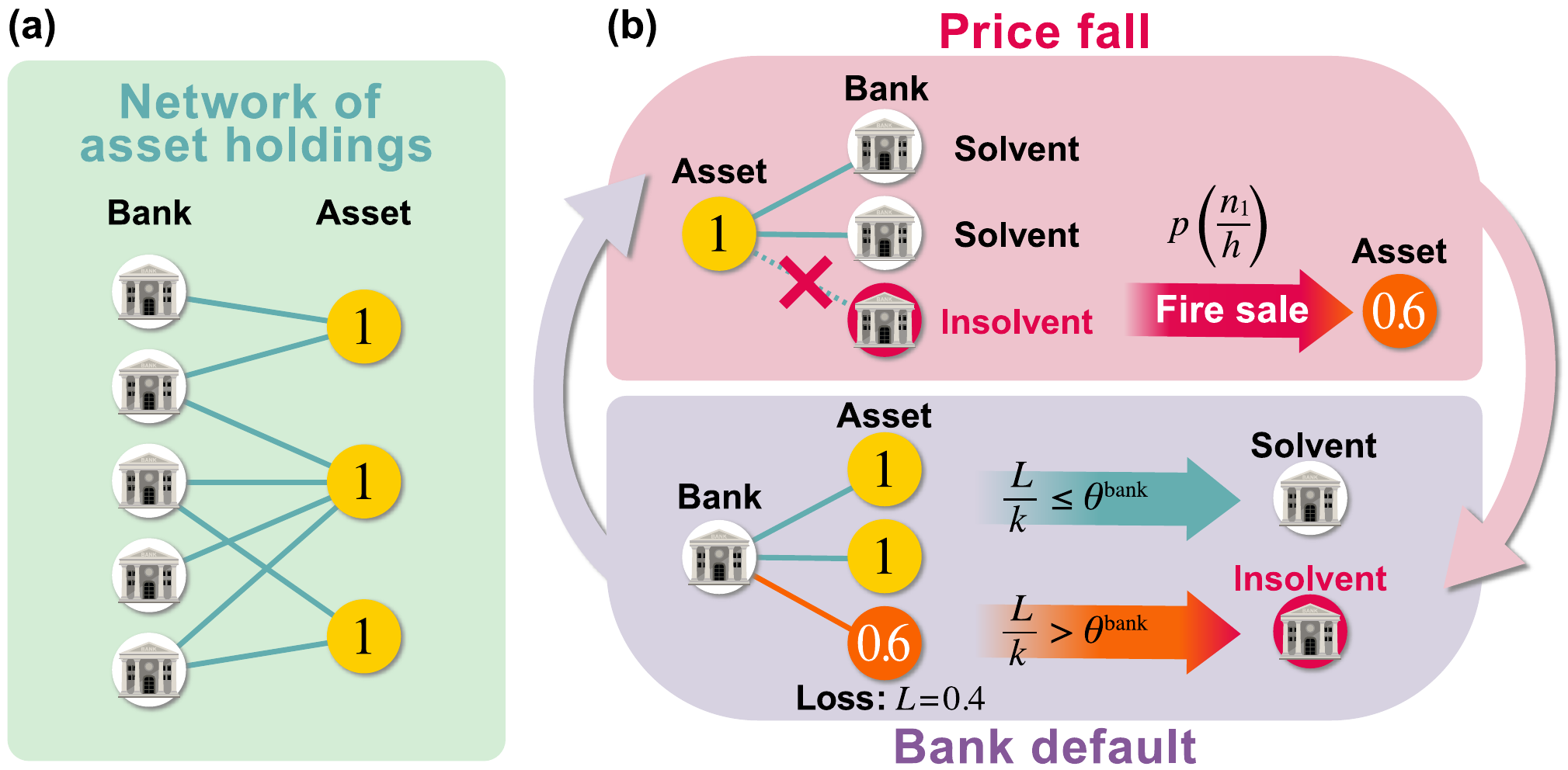}
    \caption{Schematic of fire sales through overlapping portfolios. (a) Asset-bank bipartite network. The asset prices are assumed to be $1$ in the initial state. (b) Cascade of fire sales initiated by a bank failure. The asset held by a defaulted bank will be liquidated and its price falls to 0.6 (top). Then, the price fall could cause other banks to fail due to capital loss (bottom), triggering further asset liquidations.}  
    \label{fig:schematic_bipartite}
\end{figure*}

In this section, we construct a model of financial fire sales as a dynamical process on bipartite networks, where \emph{asset nodes} take continuous states (i.e., price levels) while \emph{bank nodes} take binary states $\{0,1\}$ (i.e., solvent or insolvent). 
The asset nodes and bank nodes are connected with each other by undirected and unweighted edges, while there is no edge within the same group of nodes, forming a bipartite network.
The structure of the bipartite network represents the asset holdings, or portfolios, of banks (Fig.~\ref{fig:schematic_bipartite}a). 
For simplicity, the initial price of each asset is set to $1$ throughout the analysis.

The dynamic of contagion works as follows: A bank defaults when the value of its assets falls below a given threshold. In turn, the value of an asset depends on how many of the banks connected to the asset have defaulted (the higher the fraction of defaults, the lower the value of the asset).
We will now describe in details the model for the case of continuous-price and its multistate approximation.

\subsection{Continuous-price model}

\begin{figure}[tb]
\centering
    \includegraphics[width=8.6cm]{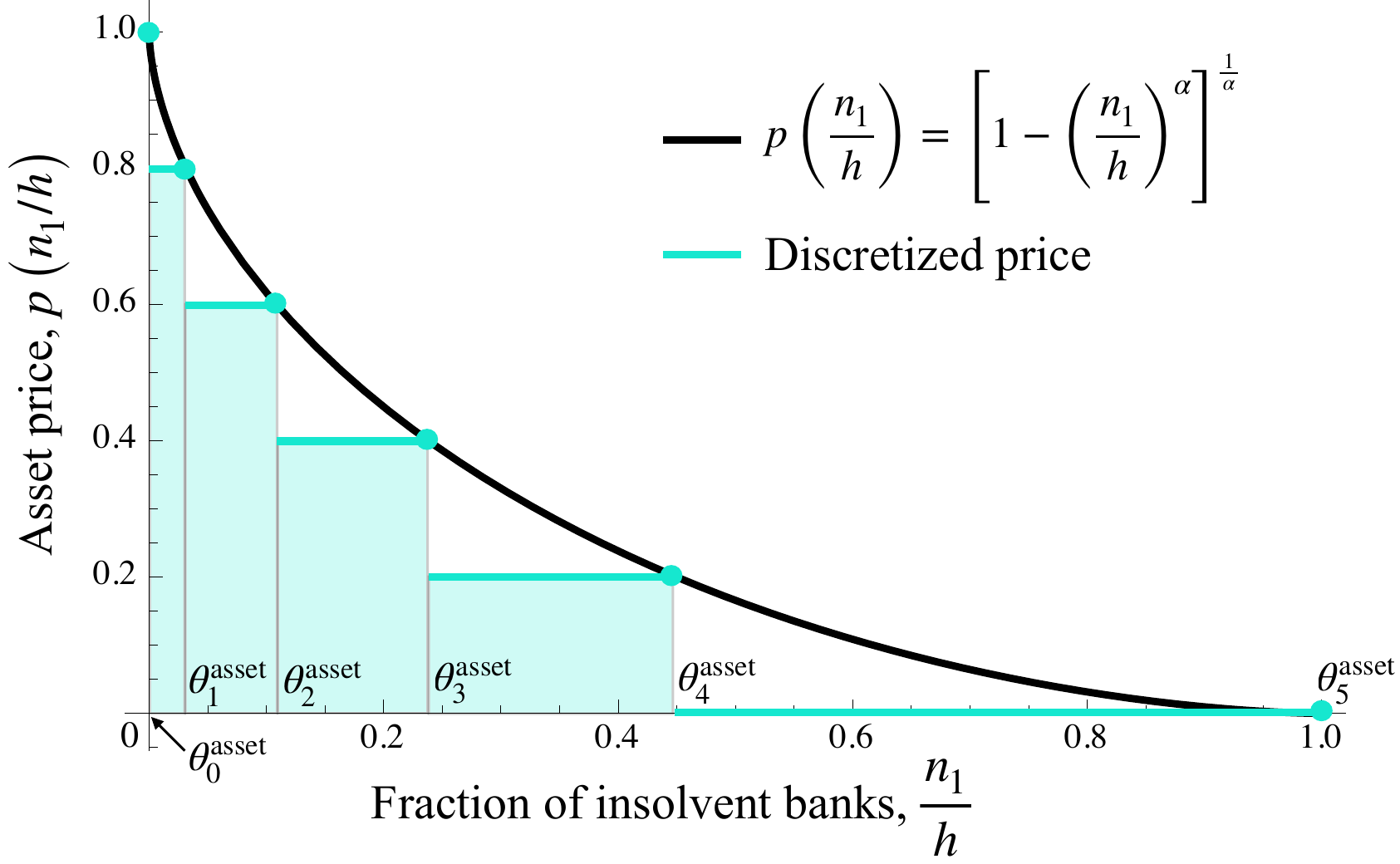}% 
    \caption{Discretization of asset prices. $n=6$ and $\alpha=0.6$.}
        \label{fig:schematic_price}
\end{figure}

 We first describe a continuous-price model in which asset prices are given by real values in $[0,1]$. 
 To specify the correspondence between asset prices and the fraction of defaulted banks, we consider a simple continuous price function of the form~\cite{Cifuentes2005JEEA}: 
\begin{align}
p\left(\frac{n_1}{h}\right) = \left[1-\left( \frac{n_1}{h}\right)^\alpha\right]^\frac{1}{\alpha},\;\; \alpha>0,
\label{eq:price_func}
\end{align}
where $h$ denotes the degree of an asset node, and $n_1$ is the number of bank nodes in state~1 (i.e., insolvent banks).
The price function \eqref{eq:price_func} captures the responsiveness of asset prices to liquidation events; the more banks fail, the lower the price of the assets held by the failed banks since those assets would be liquidated (i.e., sold in the market) (Fig.~\ref{fig:schematic_bipartite}b, \emph{top}). Eq.~\eqref{eq:price_func} has some desirable properties: $p^\prime<0$, $p(0)=1$ and $p(1)=0$ (Fig.~\ref{fig:schematic_price}). That is, the price of an asset is $1$ if there are no failed banks and $0$ if all banks that hold the asset fail.
The elasticity of asset price to the fraction of failed banks, defined by $-d\log p/d\log(n_1/h)$, is given by $x^\alpha/(1-x^\alpha)$, where $x\equiv n_1/h$. We assume $\alpha<1$ to focus on situations in which thresholds are skewed toward the origin, since otherwise the discretization approach would not work well as explained in the previous section. 

The total loss of bank $b$ due to price falls, denoted by $L_b$, is then given by
 \begin{align}
     L_b = \sum_a A_{ba}(1-p_a),
 \end{align}
 where $A = (A_{ba})$ denotes the adjacency matrix: $A_{ba}=1$ if bank $b$ holds asset $a$, and $A_{ba}=0$ otherwise. $p_a\in[0,1]$ is the price of asset $a$. Given the loss $L_b$, the state of bank $b$ shifts according to the following threshold rule:
 \begin{align}
   \mathcal{F}(0\to 1) &= 
   \begin{cases}
  1 & \text{if }\;\;  \frac{L_b}{k} > \theta^{\rm bank},\\
  0 & \text{otherwise}, 
    \end{cases}
\label{eq:threshold_condition_bank_continuous}
\end{align}
where $\theta^{\rm bank}\in (0,1)$ denotes the threshold of the fractional loss above which the bank goes bankrupt (Fig.~\ref{fig:schematic_bipartite}b, \emph{bottom})~\cite{GaiKapadia2010,Cont2013,caccioli2018review}.  The threshold $\theta^{\rm bank}$ can also be interpreted as the capital-asset ratio in the balance sheet of a bank, where a bank is forced to sell assets if the total loss $L_b$ exceeds the amount of capital $k\theta^{\rm bank}$~\cite{Cifuentes2005JEEA,caccioli2014JBF,Caccioli2015,caccioli2018review}. Since we assume that there is no possibility of recovery in the state of banks~\cite{Upper2011,Cont2013,amini2016resilience}, we have $\mathcal{F}(1\to 1)=1$.

\subsection{Multistate contagion on bipartite networks}

Now we develop a model of multistate contagion as an approximation to the continuous-price model.
Let $k$ denote the degree of a bank node, and let $\ell\in\{0,\ldots, n-1\}$ denote a state of an asset, where $\ell/(n-1)$ represents the size of price fall for the assets in state $\ell$.  
The prices of the assets held by a bank are encoded in vector $\mathbf{m}=(m_0,m_1,\ldots,m_{n-1})^\top$, where $m_\ell$ denotes the number of assets in state $\ell$, and $|\mathbf{m}|=\sum_{\ell=0}^{n-1}m_\ell=k$. The states of banks holding an asset are encoded in vector $\mathbf{n}=(n_0,n_1)^\top$, where  $n_0$ and $n_1$ denote the numbers of solvent and insolvent banks, respectively, and $|\mathbf{n}|=n_0+n_1=h$.

\subsubsection{Bank nodes}
A bank node takes one of the binary states $0$ (solvent) or $1$ (insolvent), depending on the prices of the assets that the bank has. The state-transition rate for bank nodes is given by
\begin{align}
   F_\mathbf{m}(0\to 1) &= 
   \begin{cases}
  1 & \text{if }\;\;  \frac{L(\mathbf{m})}{k} > \theta^{\rm bank},\\
  0 & \text{otherwise}, 
    \end{cases}
\label{eq:threshold_condition_bank}\\
\text{where }\;\;\;    L(\mathbf{m}) &= \frac{\sum_{\ell=1}^{n-1} \ell m_\ell}{n-1}\in[0,k].
\end{align} 
$L(\mathbf{m})$ represents the total loss that a bank incurs due to a decline in asset prices. The threshold condition~\eqref{eq:threshold_condition_bank} corresponds to Eq.~\eqref{eq:threshold_condition_bank_continuous} in the continuous-price model. 
Note that the loss from an asset price, $\ell/(n-1)$, takes (discrete) values between $0$ and $1$, so that $0\leq L(\mathbf{m})\leq k$ and $0\leq L(\mathbf{m})/k \leq 1$. 

\subsubsection{Asset nodes}

The state-transition rate for asset nodes facing a neighbor profile $\mathbf{n}$ is given by
\begin{equation}
    G_\mathbf{n}(\ell\rightarrow \ell') =
    \begin{cases}
        1 &\text{if }\; \theta_{\ell'-1}^{\rm asset} < P^{\rm asset}(\mathbf{n})\le \theta_{\ell'}^{\rm asset} \\
        0 & \text{otherwise}
    \end{cases},
    \label{eq:threshold_condition_asset}
\end{equation}
for $1\le \ell' \le n-1$, where $P^{\rm asset}(\mathbf{n})=n_1/h$ is the fraction of insolvent banks.
$\theta_{\ell'-1}^{\rm asset}$ and $\theta_{\ell'}^{\rm asset}$ are the threshold values between which the asset is in state $\ell'$. The threshold condition \eqref{eq:threshold_condition_asset} indicates that an asset price changes in a discrete manner in accordance with the fraction of insolvent banks among all the asset holders.
For $\ell'=0$, we have
\begin{equation}
    G_\mathbf{n}(\ell\rightarrow 0) =
    \begin{cases}
        1 &\text{if }\; P^{\rm asset}(\mathbf{n})\leq \theta_{0}^{\rm asset} \\
        0 & \text{otherwise}
    \end{cases}
    \label{eq:response_G}
\end{equation}
This suggests that when $\theta_{0}^{\rm asset}=0$, an asset price will remain unchanged if and only if there are no failed banks that hold the asset. Note that when $n=2$, the model reduces to a model of binary-state contagion on bipartite networks~\cite{lux2016jedc,dodds2018simple}.

By implementing a discretization such that the price of assets in state-$\ell$ is equal to $1-\ell/(n-1)$, the threshold values are given by 
\begin{equation}
    \theta_\ell^{\rm asset} = \left[1-\left(1-\frac{\ell}{n-1}\right)^\alpha\right]^\frac{1}{\alpha}, \; \ell=0, \ldots,n-1.
    \label{eq:theta_asset}
\end{equation}
Note that $\theta_0^{\rm asset}=0$ and $\theta_{n-1}^{\rm asset}=1$ (Fig.~\ref{fig:schematic_price}). Clearly, the set of threshold values in the limit of $n\to \infty$ leads to $\{\theta: \theta\in[0,1]\}$, in which case the asset prices are determined by the continuous price function~\eqref{eq:price_func}. Thus, the continuous-price model is also interpreted as a continuous-threshold model as long as the price function is continuous.

It should be noted that due to the discretization of thresholds, the multistate cascade process only provides an approximation to the ``true'' dynamics that would be observed if the asset price would follow the original price function~\eqref{eq:price_func}. Nonetheless, we will show in section~\ref{sec:result_convergence} that the simulated cascade dynamics obtained with the continuous price function can be well replicated in the discrete multistate model for a moderately large $n$.

\subsection{Approximate master equations}
Let $x^i_{k,\mathbf{m}}(t)$ denote the fraction of $k$-degree banks that are in state $i$ and face asset-price profile $\mathbf{m}$ at time $t$. For each degree $k$, we have $\sum_{l=0}^{n-1}m_\ell=k$.
The expected fraction of banks in state $i$ at time $t$ is given by
\begin{align}
    \rho_i^{\rm bank}(t) = \left\langle\sum_{|\mathbf{m}|=k} x^i_{k, \mathbf{m}}(t)\right\rangle\subrangle{k},
    \label{eq:rho_i_bank}
\end{align}
where $\langle\cdot\rangle_k$ denotes the average over $k$ with degree distribution $p_k^{\rm bank}$. 

For asset nodes, we define $y^\ell_{h,\mathbf{n}}(t)$ as the fraction of $h$-degree assets that are in state $\ell$ and face asset holders' profile $\mathbf{n}$. 
The expected fraction of assets in state $\ell$ at time $t$ leads to
\begin{equation}
    \rho_{\ell}^{\rm asset} = \left\langle \sum_{|\mathbf{n}|=h} y^\ell_{h, \mathbf{n}}(t) \right\rangle\subrangle{h},
\end{equation}
where $\langle\cdot\rangle_h$ denotes the average over $h$ with degree distribution $p^{\rm asset}_h$. 

By employing the AME formalism, the dynamics of $x^i_{k,\mathbf{m}}$ and $y^l_{h,\mathbf{n}}$ are respectively described by the following differential equations:
\begin{eqnarray}
    \frac{dx^i_{k,\mathbf{m}}}{dt} &=& - \delta_{i0}F_\mathbf{m}(0\rightarrow 1) x^0_{k,\mathbf{m}}
    + \delta_{i1}F_\mathbf{m}(0\rightarrow 1) x^0_{k,\mathbf{m}} \nonumber\\
    && - \sum_{\ell=0}^{n-1}\sum_{\ell'>\ell} m_\ell\beta^i(\ell\rightarrow \ell') x^i_{k,\mathbf{m}} \nonumber \\
     &&+ \sum_{\ell=0}^{n-1}\sum_{\ell'<\ell} (m_{\ell'}+1)\beta^i(\ell'\rightarrow \ell) x^i_{k,\mathbf{m}-\mathbf{e}_\ell+\mathbf{e}_{\ell'}}, 
    \label{eq:AME_finanical_x}\\
    \frac{dy^\ell_{h,\mathbf{n}}}{dt} &=& - \sum_{\ell'>\ell}G_\mathbf{n}(\ell\rightarrow \ell') y^\ell_{h,\mathbf{n}}
    + \sum_{\ell'<\ell}G_\mathbf{n}(\ell'\rightarrow \ell) y^{\ell'}_{h,\mathbf{n}} \nonumber\\
    && - n_0\gamma^\ell(0\rightarrow 1) y^\ell_{h,\mathbf{n}} \nonumber \\
    && + (n_{0}+1)\gamma^\ell(0\rightarrow 1) y^\ell_{h,\mathbf{n}-\mathbf{e}_1+\mathbf{e}_{0}},
\label{eq:AME_finanical_y}
\end{eqnarray}
where $\delta_{ij}$ is the Kronecker delta.
The link transition rates $\beta^i(\ell\rightarrow \ell')$ and $\gamma^\ell(i\rightarrow j)$ are given by
\begin{align}
\beta^i(\ell\rightarrow \ell') &= \frac{
\langle \sum_{|\mathbf{n}|=h} n_i y^\ell_{h,\mathbf{n}}(t) G_\mathbf{n}(\ell\rightarrow \ell') \rangle_h
}{
\langle \sum_{|\mathbf{n}|=h} n_i y^\ell_{h,\mathbf{n}}(t) \rangle_h
},\\
\gamma^\ell(i\rightarrow j) &= \frac{
\langle \sum_{|\mathbf{m}|=k} m_\ell x^i_{k,\mathbf{m}}(t) F_\mathbf{m}(i\rightarrow j) \rangle_k
}{
\langle \sum_{|\mathbf{m}|=k} m_\ell x^i_{k,\mathbf{m}}(t) \rangle_k
}.
\end{align}
Eqs.~\eqref{eq:AME_finanical_x} and \eqref{eq:AME_finanical_y} are obtained in the same way as the standard AMEs (Eq.~\ref{eq:AME_eq}). Eq.~\eqref{eq:AME_finanical_x} describes the dynamics of bank states while Eq.~\eqref{eq:AME_finanical_y} represents the dynamics of the share of state-$\ell$ assets. Since there is no edge within the same group, the states of neighbors for bank nodes (resp. asset nodes) correspond to the price of assets that the bank has (resp. the solvency of the asset-holding banks). $\beta^i(\ell\rightarrow \ell')$ thus denotes the probability of an asset held by a state-$i$ bank changing its state from $\ell$ to $\ell'$. Analogously, $\gamma^\ell(i\rightarrow j)$ denotes the chance that the state of a bank holding state-$\ell$ assets shifts from $i$ to $j$. Combining Eqs.~\eqref{eq:AME_finanical_x} and \eqref{eq:AME_finanical_y}, the total number of differential equations to be solved is  $2\sum_{k=0}^{k_{\max}}\binom{k+1}{k} + n\sum_{h=0}^{h_{\max}}\binom{n+h-1}{h}$.

\begin{figure*}[tb]
\centering
    \includegraphics[width=17.7cm]{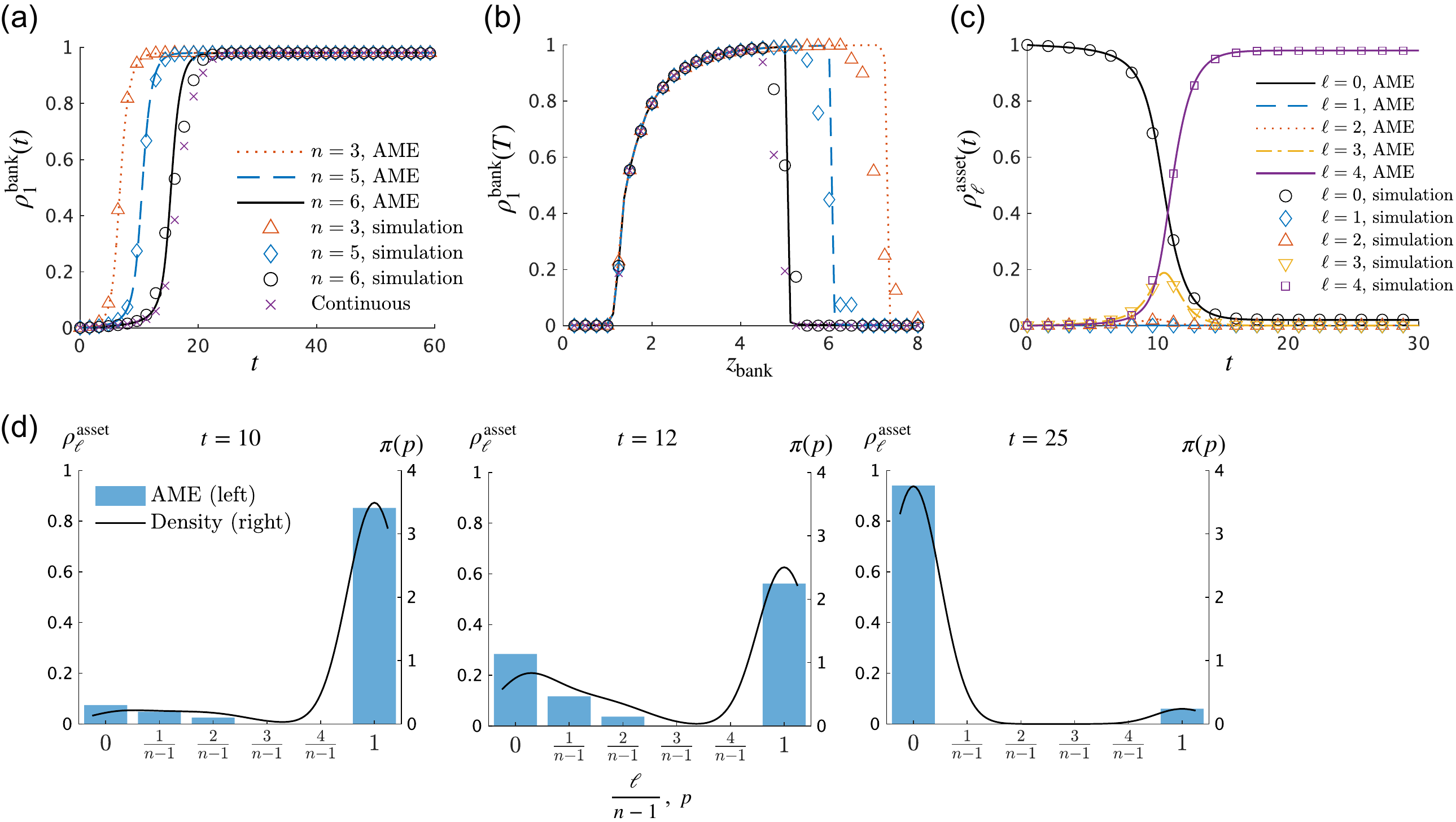}
    \caption{Comparison between theory and simulation. (a) Dynamical paths of $\rho_1^{\rm bank}(t)$ obtained by the AME method (lines) and simulation (symbols). We set $z_{\rm bank}=4$, $N=5000$, and $M=5000$. (b) Steady-state fraction of insolvent banks against the bank degree. (c) The average fraction of assets in each state. 
    (d) Distribution of asset prices calculated by the AME (blue bar) and numerical simulations in the continuous-price model (black line). Density of the continuous prices is obtained by the kernel density estimation with bandwidth $0.1$.   
    Unless otherwise noted, we set $k_{\max} = \min\{k': \sum_{k=0}^{k'} z_{\rm bank}^k e^{-z}/k! \ge 0.99\}$, $\rho_1^{\rm bank}(0)=10^{-3}$, $\alpha=0.6$, $\theta^{\rm bank}=0.2$, $N=10^4$, $M=10^4$, $T=60$ and $dt=0.01$. The average is taken over 40 runs.
    }
\label{fig:AME_comparison_benchmark} 
\end{figure*}

\section{Results}

\subsection{Convergence to the continuous-state equilibrium}\label{sec:result_convergence}

Now we calculate the average size of default cascades $\rho_1^{\rm bank}(t)$ based on Eqs.~\eqref{eq:AME_finanical_x} and \eqref{eq:AME_finanical_y}, assuming that banks and asses are connected uniformly at random; bank degrees follow a Poisson distribution with mean $z_{\rm bank}$, and a bank with degree $k$ is connected to an asset with probability $k/M$, where $M$ is the total number of asset nodes. 
The number of edges is given by $Nz_{\rm bank}$, where $N$ is the number of banks. Since the presence of multiedges can be ignored as long as $z_{\rm bank}/M$ is sufficiently small, the degree distribution for assets is also Poissonian with mean $Nz_{\rm bank}/M$.

Fig~\ref{fig:AME_comparison_benchmark}a shows that simulated paths of the average fraction of insolvent banks are well replicated by the corresponding AME solutions. 
We find that a global cascade can occur only in a certain range of network connectivity captured by $z_{\rm bank}$ (Fig.~\ref{fig:AME_comparison_benchmark}b), which is consistent with many studies on binary-state cascades~\cite{Watts2002,GaiKapadia2010,Brummitt2015PRE,kobayashi2015trend,caccioli2018review}. Note that the cascade region gradually shrinks as the number of states $n$ increases since the discretized threshold values $\{\theta_{\ell}^{\rm asset}\}$ tend to rise on average (Fig.~\ref{fig:schematic_price}). In particular, when $n=6$, the the steady-state fraction $\rho_1^{\rm bank}(T)$ calculated by the AME (black solid in Fig.~\ref{fig:AME_comparison_benchmark}a and b) well matches the corresponding steady state in the continuous-state model (purple cross), indicating that $\rho_1^{\rm bank}(t)$ converges to the continuous equilibrium as $n$ increases. That is, the solution to the continuous-state model can be accurately replicated with a finite set of discretized states as long as $n$ is moderately large. Our multistate model thus provides an analytical foundation to the previously proposed numerical models of financial fire sales with continuous prices~\cite{Huang2013SciRep,caccioli2014JBF, caccioli2018review, vodenska2021systemic}.

In addition to the average fraction of defaulted banks, the share of assets in each state and even the distribution of continuous asset prices are approximated by the AME when the number of states is moderately large (Fig.~\ref{fig:AME_comparison_benchmark}c and d). 
The figures show that the price distribution drastically changes from $\rho_0^{\rm asset}\approx 1$ to $\rho_{n-1}^{\rm asset}\approx 1$ around $t = 12$, after which most of the asset prices drop to $0$.

\begin{figure}[thb]
    \centering
    \includegraphics[width=8.6cm]{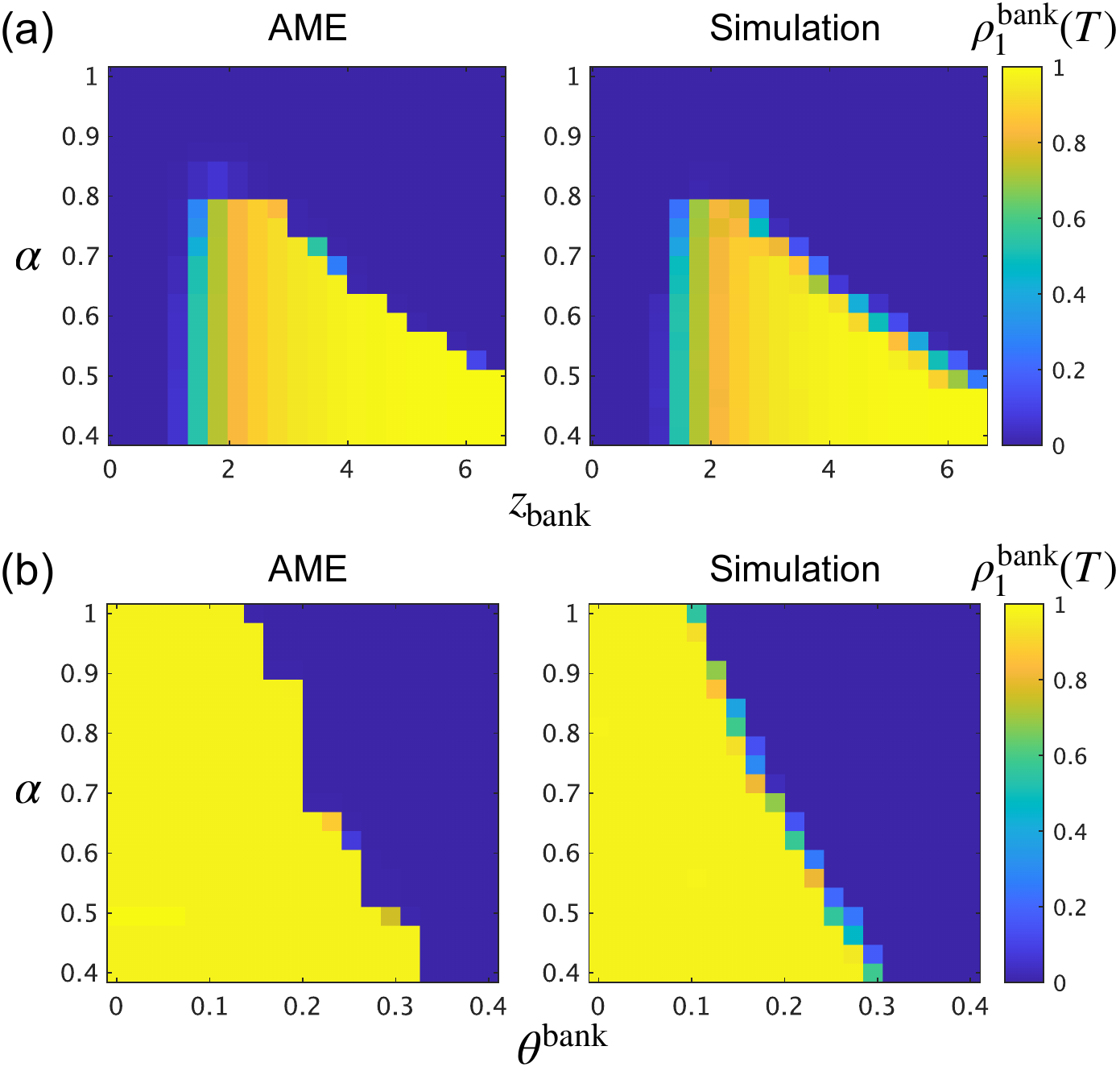}
    \caption{Cascade region obtained by the AMEs and simulated continuous-price model. (a) $\alpha$ vs. $z_{\rm bank}$, and (b) $\alpha$ vs. $\theta^{\rm bank}$. We set $n=6$ for the AME method. 
    See Fig.~\ref{fig:AME_comparison_benchmark} for the other parameter values.
    }
\label{fig:AME_cascade_region}
\end{figure}

Fig.~\ref{fig:AME_cascade_region} presents a comparison of the cascade regions obtained by theory and simulation under different parameter combinations. First, the larger the elasticity parameter $\alpha$, the smaller the cascade region.  Generally, the average of discretized threshold values increases as $\alpha$ rises, thereby shrinking the cascade region (Eq.~\ref{eq:theta_asset}). Second, the effect of network connectivity $z_{\rm bank}$ is twofold, as is well known in the literature on threshold cascades; a network needs to be moderately sparse for a global cascade to occur since there must be routes through which influence spreads and, at the same, the average degree needs to be sufficiently low so that influence from a neighbor would not be diluted~\cite{Watts2002,GaiKapadia2010}.

\subsection{Preventing cascades through policy intervention}

In this section, we explore how cascades of fire sales could be avoided by intervention. Here, we consider policy interventions in which the government/financial authority may operate in the market to reduce the probability of asset prices declining or to prevent banks from going bankrupt. In practice, the former type of policies has been implemented extensively through massive asset-purchasing programs introduced in response to the global financial crisis of 2007--2009~\cite{cecchetti2009crisis,taylor2010getting,mishkin2011over}. Along with this, defaulting banks have been occasionally rescued by the governments' capital injection (e.g., purchasing the banks' stocks) over the past decades~\cite{kollmann2012fiscal,brei2013rescue,hoshi2015will}. Within our analytical framework, each of these policies may be represented as an alteration to the state-transition probabilities in $G_\mathbf{n}$ or $F_\mathbf{m}$.  

To analyze the effect of an ``asset-purchasing policy," we introduce a probability $q_{\rm asset}$ that an asset price does not change even if a sufficient fraction of its asset holders newly default. 
That is, with probability $1-q_{\rm asset}$, the state of an asset shifts as is specified by Eq.~\eqref{eq:threshold_condition_asset}, and with probability $q_{\rm asset}$, the asset's state is kept unchanged regardless of the state of the asset holders. 
Analogously, we can also consider another policy tool that could prevent banks from being defaulted. Such a ``bank-rescue policy" is represented by a probability $q_{\rm bank}$ that the state of a bank is kept unchanged regardless of the asset prices that the bank has (see Appendix~\ref{sec:SI_threshold_policy} for the specification of the modified transition probabilities). Clearly, setting $q_{\rm asset}=q_{\rm bank}=0$ recovers the baseline model in which there is no policy intervention.

\begin{figure}[tb]
\centering
    \includegraphics[width=6.3cm]{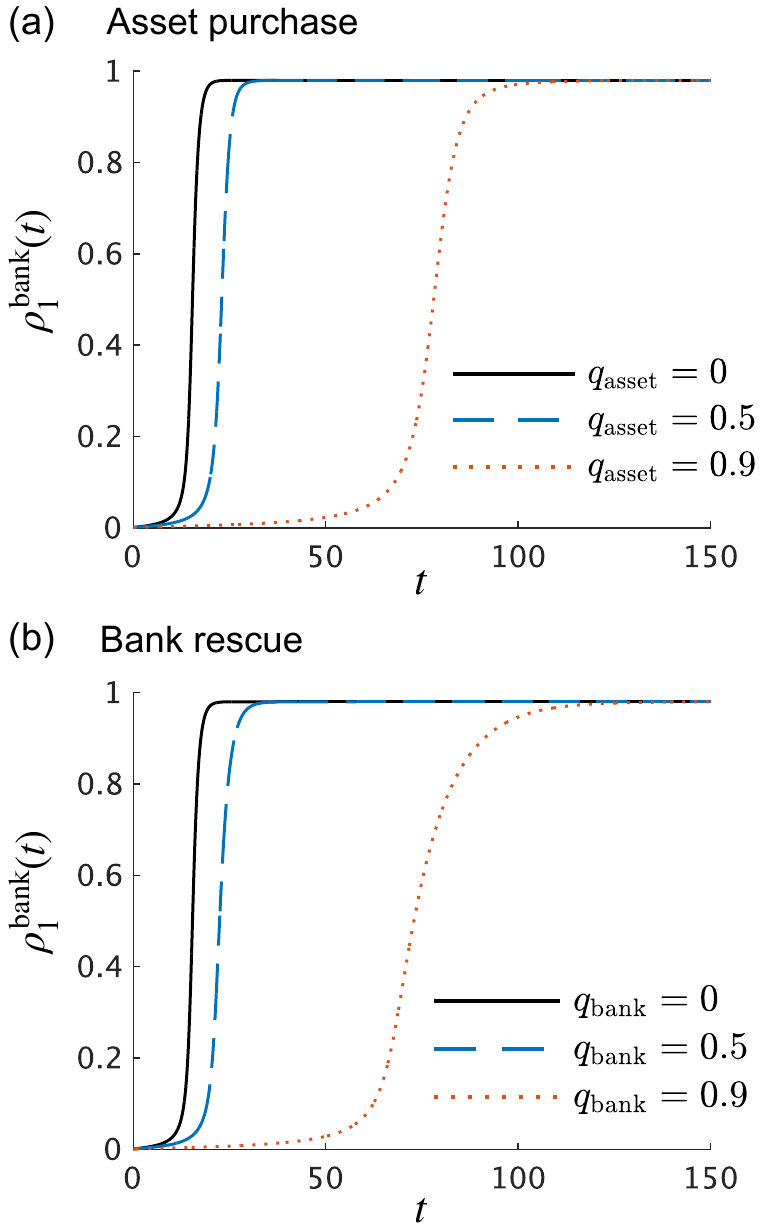}
    \caption{Effect of policy intervention on the fraction of defaulted banks. (a) Asset-purchasing policy and (b) bank-rescue policy. $q_{\rm asset}$ and $q_{\rm bank}$ respectively represent the probabilities that the price of an asset and the state of a bank are kept unchanged due to policy intervention.}
\label{fig:AME_intervention}
\end{figure}

We find that both policies have a similar effect on cascades (Fig.~\ref{fig:AME_intervention}); the possibility of policy intervention delays the onset of fire sales while the terminal state attained around $t\approx 100$ is unaffected (i.e., $\rho_1^{\rm bank}\approx 1$).
In the process of propagation, the cascade region naturally shrinks as $q_{\rm asset}$ increases at a given point in time (Fig.~\ref{fig:ame_intervention_asset}). This suggests that a probabilistic policy intervention may delay the propagation of fire sales, but eventually the undesirable steady state will be attained.

\subsection{Application to the ETF market}

\begin{figure}[tb]
 \centering
    \includegraphics[width=8.6cm]{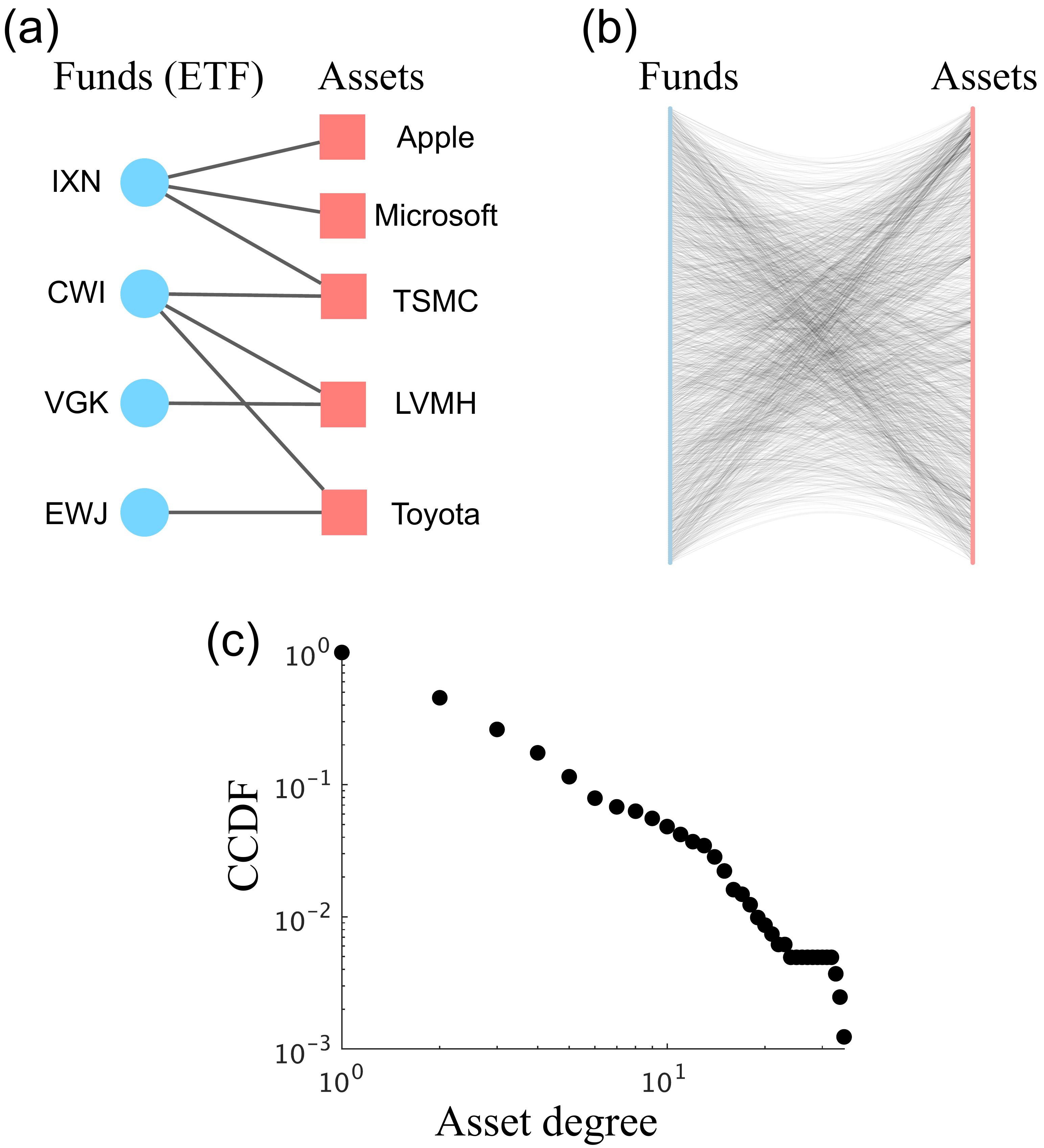}
    \caption{Bipartite network of ETF portfolios. (a) Schematic of the bipartite network between funds and assets. For instance, the ETF labeled ``IXN'' holds the stocks of Apple, Microsoft, and TSMC.
    (b) Visualization of the actual ETF network, where the number of funds $N$ is $212$ (light blue), and the number of assets $M$ is $810$ (pale red). (c) Complementary cumulative distribution function (CCDF) for the asset degrees. }  
\label{fig:etf_schematic}
\end{figure}

\begin{figure}[t]
\centering
    \includegraphics[width=8.6cm]{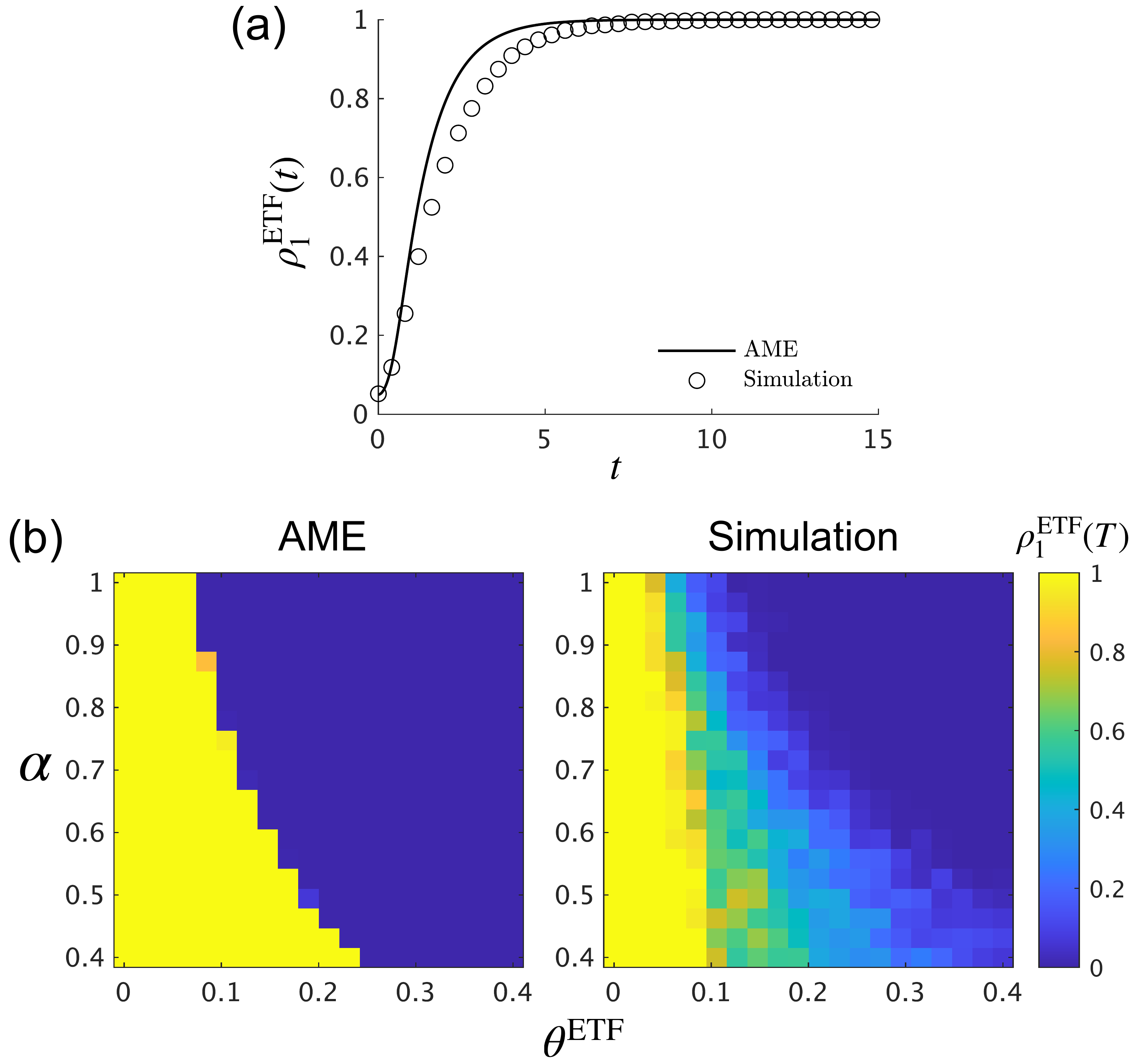}
    \caption{Comparison between theory and simulation for the US ETF portfolios. $\rho_1^{\rm ETF}$ and $\theta^{\rm ETF}$ correspond to $\rho_1^{\rm bank}$ and $\theta^{\rm bank}$ in the baseline model, respectively. Simulated results for a continuous-price model are compared with the corresponding AME solutions with $n=6$. (a) Dynamical path of the fraction of collapsed ETFs and (b) the cascade region. The initial seed fractions $\rho_1^{\rm ETF}(0)$ are set at $0.05$ and $1/N$ in panels a and b, respectively. In panel a, we set $\alpha= 0.6$ and $\theta^{\rm ETF}= 0.04$. In panel b, the color denotes the average over 40 runs for $T=20$.}
    \label{fig:ETF_path_heatmap_main}
\end{figure}

In this section we examine the possibility of a global cascade occurring in a real-world financial network. 
We use data on the asset portfolios of exchange-traded funds (ETFs) listed in exchanges in the US (e.g., NASDAQ, NYSE Arca). 
The ETF data are collected from the website of Monex Inc.~\cite{Monex}.
We consider a bipartite network in which there are two types of nodes, ETFs (i.e., fund nodes) and stocks (i.e., asset nodes). A fund node and an asset node are connected by an undirected and unweighted edge if the asset is held by the ETF (Fig.~\ref{fig:etf_schematic}a). The network data is available in Github at~\cite{Kawamoto_Github}. 
Due to the limited availability of portfolio information, we use the top 10 assets for each ETF, in terms of the share in the portfolio, in constructing the bipartite network. To capture systemic risk in the ETF market, rather than the risk of each individual asset, we combine several individual stocks that apparently belong to the same group (e.g., BHP Group Ltd and BHP Group Plc). 
 Therefore, each fund node is connected to at most ten asset nodes (i.e., $k\leq 10$ for all fund nodes). 
 We focus on the largest connected component, where the numbers of the fund nodes and the asset nodes are $N=212$ and $M=810$, respectively (Fig.~\ref{fig:etf_schematic}b).
 We find that the asset degrees are quite heterogeneous and the degree distribution is heavy-tailed (Fig.~\ref{fig:etf_schematic}c).

In the ETF network, fund nodes correspond to bank nodes in the baseline model, so we assume that an ETF will be liquidated (i.e., collapsed) if its loss exceeds a certain fraction $\theta^{\rm ETF}$, in the same way that a bank defaults according to the threshold condition \eqref{eq:threshold_condition_bank}.
Fig.~\ref{fig:ETF_path_heatmap_main} compares the solution of the AME and the corresponding simulation results in the continuous-price model. In the AME method, we use the actual degree distributions of the ETF network in calculating the fraction of collapsed ETFs, denoted by $\rho_1^{\rm ETF}$. The dynamical path of $\rho_1^{\rm ETF}(t)$ suggested by the AME method well matches the simulated values (Fig. \ref{fig:ETF_path_heatmap_main}a) while the network structure is not random as assumed in the baseline model.

% It should be noted that an advantage of a bipartite network in using the AME method is that there is no loop by construction; it is well known that the presence of local loops, which are often seen in many empirical social networks, would generally undermine the accuracy of approximation methods such as message-passing methods~\cite{cantwell2019message}. 
% In bipartite networks, the accuracy of an approximation method will not be deteriorated by local repercussions of influences, which can be a non-negligible problem especially for multistate, as well as binary, contagion models~\cite{Melnik2013}. In this sense, the bipartite structure of the ETF portfolios may be advantageous for the accuracy of the AME method.

On the other hand, since the AME method is developed for sufficiently large networks, there are necessarily some discrepancy between the AME solution and simulation results stemming from the finite-size effect. The finite-size effect is also visible in Fig.~\ref{fig:ETF_path_heatmap_main}b, which shows the cascade region for different values of $\alpha$ and $\theta^{\rm ETF}$. Nevertheless, simulated cascades in the continuous-price model are well predicted by the discrete-state AME method with $n=6$.

\section{Conclusion}

We develop an analytical framework for the continuous-state model of complex contagion which can be extended to describe the dynamics of financial fire sales with continuous prices. While the discretized model involves a limited number of states, its AME solution well matches the continuous-state equilibrium as long as the number of possible states is moderately large. The analytical tractability allows us to examine various properties, such as the size of cascade region and the effects of policy interventions, without running massive numerical simulations.

There are still some issues to be addressed in future research. First, while the model is based on a system of differential equations, the cascade condition is not obtained in an analytical manner as is done in many binary-state models~\cite{Watts2002,Gleeson2007,Gleeson2008,Brummitt2015PRE,kobayashi2015trend}. We in fact found that the maximum eigenvalue of the Jacobian of the differential equations is not necessarily informative in predicting the cascade region. Arguably, this is because the system is highly nonlinear so that a local linearization around the initial point may not accurately capture the stability of the system. Second, since our method describes the collective dynamics that would materialize on average, it is not possible to theoretically identify important nodes in the network. In practice, however, there is a substantial need for identifying systemically important banks in terms of their possible impact on financial cascades. Finally, the idea that a discrete multistate contagion model can provide an approximation of continuous-state contagion could also be utilized in other contexts, such as opinion dynamics and cascades of load~\cite{Brummitt2012_PNAS,perra2019opinion,baumann2020modeling}.
We hope that our work will stimulate further research in these directions.

\section*{Acknowledgments}
T.\ K.\ acknowledges financial support from JSPS KAKENHI\ 19H01506, 20H05633 and 	22H00827. T.\ O.\ acknowledges financial support from JSPS KAKENHI\ 19K14618 and 19H01506. We would like to thank Tatsuro Kawamoto and Yoshitaka Ogisu for constructing the ETF data set, and James Gleeson and Takehisa Hasegawa for valuable comments.

\appendix

\section{Discretization of threshold values}
\label{sec:threshold_discretization}

\begin{figure*}[tb]
    \centering
    \includegraphics[width=17cm]{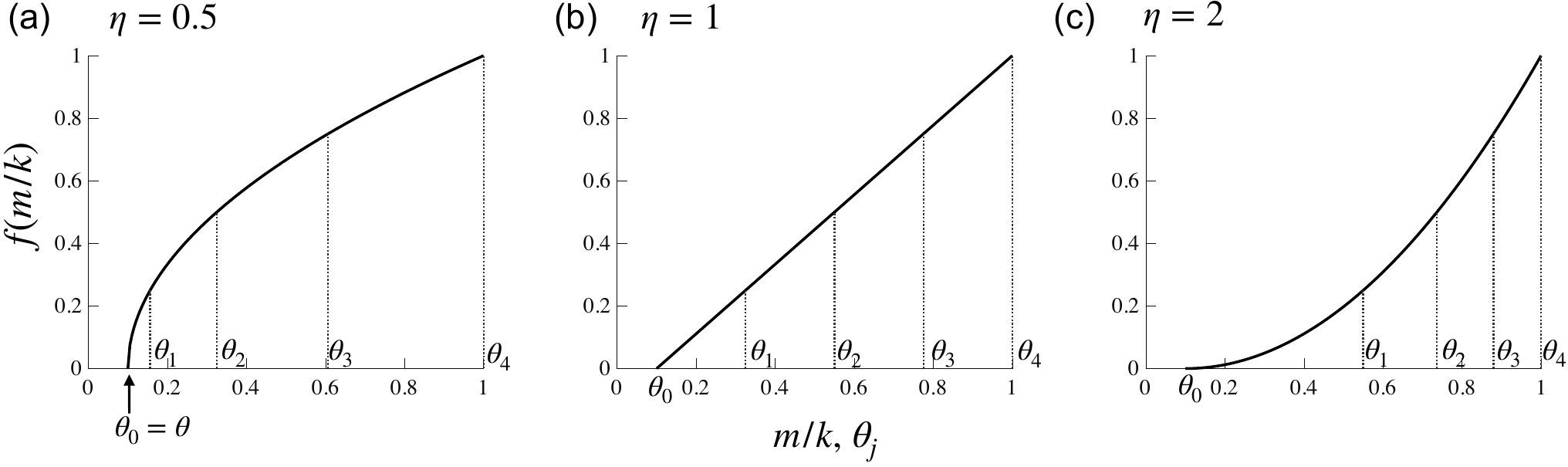}
    \caption{
    Specification of threshold values $\theta_j$ for $n=5$. The responsiveness in the continuous response function is given by $f(m/k) = \left(\frac{m/k-\theta}{1-\theta}\right)^\eta$ (Eq.~\ref{eq:F_continuous_watts_nonlin}). For given $\eta$ and $\theta$, discretized thresholds are determined by Eq.~\eqref{eq:discretized_threshold_SI}, where each interval size for the responsiveness is given by $f(\theta_{j})-f(\theta_{j-1})=\frac{1}{n-1}$ and is common for all $j=1,\ldots,n-1$. $\theta(=\theta_0 )$ is set at $0.1$, while the values of $\eta$ are (a) 0.5, (b) 1, and (c) 2. The distribution of $\{\theta_j\}$ is skewed to the right as $\eta$ rises, meaning that the response to the neighbors' states becomes more conservative. }
    \label{fig:schematic_response_SI}
\end{figure*}

In the continuous-state threshold model with $\theta\geq 0$, the response function can be expressed as:
\begin{align}
    \mathcal{F}_{\vect{s}} =  
    \begin{cases}
   \left( \frac{m/k-\theta}{1-\theta}\right)^\eta\; {\text{ if } }\;  \frac{m}{k} > \theta  \\
         0 \;\text{ otherwise},
     \end{cases}
    \label{eq:F_continuous_watts_nonlin}
\end{align}
where $f(m/k)=\left( \frac{m/k-\theta}{1-\theta}\right)^\eta$ represents the responsiveness to the neighbors' states. Note that $f$ satisfies $f(\theta)=0$ and $f(1)=1$. $\eta\geq 0$ captures the strength parameter; when $\eta<1$, the response is ``progressive'' in that the slope of $f(m/k)$ is larger than that in the case of $\eta=1$ when $m/k$ is close to $\theta$  (Fig.~\ref{fig:schematic_response_SI}a and b), while the response becomes ``conservative'' when $\eta>1$ (Fig.~\ref{fig:schematic_response_SI}c). Note that the Watts model of binary-state cascades corresponds to the most progressive response function where $\eta = 0$. 

In the discrete-state model, the threshold values corresponding to the above response function are given by
\begin{align}
    \theta_j &= \theta +  (1-\theta)\left(\frac{j}{n-1}\right)^\frac{1}{\eta}, \notag \\
             &\;\;j=0,\ldots,n-1,\; 0<\eta<\infty,
    \label{eq:discretized_threshold_SI}
\end{align}
where $\theta$ is the threshold given in Eq.~\eqref{eq:F_continuous_watts_nonlin} and is equal to $\theta_0$.
This specification ensures that each interval size between two adjacent responsiveness values is common: $f(\theta_{j})-f(\theta_{j-1})=\frac{1}{n-1}$, for all $j=1,\ldots,n-1$.
Fig.~\ref{fig:schematic_response_SI} illustrates three cases: progressive, neutral and conservative responses.

We can also consider an alternative nonlinear response function based on the sum of neighbors' states:
\begin{align}
    \widetilde{\mathcal{F}}_{\vect{s}} =  
    \begin{cases}
   \left( \frac{\sum_j s_j/k-\theta}{1-\theta}\right)^\eta\; {\text{ if } }\;  \frac{\sum_j s_j}{k} > \theta  \\
         0 \;\text{ otherwise}.
     \end{cases}
    \label{eq:F_continuous_watts_nonlin_tilde}
\end{align}
Fig.~\ref{fig:continuous_AME_sum} reveals that the dynamical paths predicted by the AME method well match the simulated ones when $\eta<1$. This suggests that the AME method provides a good approximation when $\eta<1$, in which case threshold values are skewed to the left. When $\eta\geq 1$, the AME method still predicts that a cascade occurs while simulation results do not (Fig.~\ref{fig:continuous_AME_sum}, \emph{middle} and \emph{bottom}).

\section{Transition probabilities in the presence of policy intervention}
\label{sec:SI_threshold_policy}

In the presence of an asset-purchasing policy, the transition matrix $G_
{\mathbf{n}}$ in the baseline model (Eq.\eqref{eq:threshold_condition_asset}) is modified since the policy alters the probability of an asset price declining.
For $\ell,\ell'\geq 1$, we have
\begin{widetext}
\begin{align}
    G_\mathbf{n}(\ell\rightarrow \ell') =
    \begin{cases}
        q_{\rm asset} &\text{if }\;  \ell\notin\{\ell:\theta_{\ell-1}^{\rm asset} < P^{\rm asset}(\mathbf{n})\le \theta_{\ell}^{\rm asset}\}  \text{ and }\ell'=\ell \\
        1-q_{\rm asset} &\text{if }\;  \ell\notin\{\ell:\theta_{\ell-1}^{\rm asset} < P^{\rm asset}(\mathbf{n})\le \theta_{\ell}^{\rm asset}\}  \text{ and }
        \ell'\in\{\ell':\theta_{\ell'-1}^{\rm asset} < P^{\rm asset}(\mathbf{n})\le \theta_{\ell'}^{\rm asset}\}\\
        1 &\text{if }\;  \ell\in\{\ell:\theta_{\ell-1}^{\rm asset} < P^{\rm asset}(\mathbf{n})\le \theta_{\ell}^{\rm asset}\}  \text{ and }\ell'=\ell \\
        0 & \text{otherwise}
\end{cases}.
    \label{eq:G_policy}
\end{align}
For $\ell=0$ or $\ell'=0$, we have
\begin{align}
    G_\mathbf{n}(0\rightarrow \ell') = \begin{cases}
        q_{\rm asset} &\text{if }\;  P^{\rm asset}(\mathbf{n}) > \theta_{0}^{\rm asset}   \text{ and }\ell'= 0 \\
        1-q_{\rm asset} &\text{if }\;  \ell'\in\{\ell':\theta_{\ell'-1}^{\rm asset} < P^{\rm asset}(\mathbf{n})\le \theta_{\ell'}^{\rm asset}\} \text{ and }\ell'\geq 1\\
        0 & \text{otherwise}
\end{cases}, 
\end{align}
\begin{align}
    G_\mathbf{n}(\ell\rightarrow 0) &=
    \begin{cases}
        1 &\text{if }\;   P^{\rm asset}(\mathbf{n}) \le \theta_{0}^{\rm asset}  \\
        0 & \text{otherwise}
\end{cases}.
    \label{eq:G_policy_ell_0}
\end{align}
\end{widetext}
The probability $q_{\rm asset}$ thus represents the chance that an asset-purchasing policy is implemented, where setting $q_{\rm asset}=0$ recovers the baseline model without policy intervention.  

In the presence of a bank-rescue policy, the transition matrix $F_\mathbf{m}$ is given by
\begin{align}
   F_\mathbf{m}(0\to 1) &= 
   \begin{cases}
  1-q_{\rm bank} & \text{if }\;\;  \frac{L(\mathbf{m})}{k} > \theta^{\rm bank},\\
  0 & \text{otherwise}, 
    \end{cases} \\
   F_\mathbf{m}(0\to 0) &= 
   \begin{cases}
  q_{\rm bank} & \text{if }\;\;  \frac{L(\mathbf{m})}{k} > \theta^{\rm bank},\\
  0 & \text{otherwise},
    \end{cases}
\end{align} 
where $q_{\rm bank}$ denotes the chance that the government rescues the defaulting bank. Clearly, $q_{\rm bank}=0$ recovers the baseline model in which there is no policy intervention.

%\vspace{5cm}

%\bibliography{financial_multistage}% Produces the bibliography via BibTeX.


\clearpage

%\appendix
%%%%%%%%%% Merge with supplemental materials %%%%%%%%%%
%%%%%%%%%% Prefix a "S" to all equations, figures, tables and reset the counter %%%%%%%%%%
\setcounter{equation}{0}
\setcounter{figure}{0}
\setcounter{table}{0}
\setcounter{page}{1}
\setcounter{section}{0}

\makeatletter
\renewcommand{\theequation}{S\arabic{equation}}
\renewcommand{\thefigure}{S\arabic{figure}}
\renewcommand{\thesection}{S\arabic{section}}

\renewcommand{\bibnumfmt}[1]{[S#1]}
\renewcommand{\citenumfont}[1]{S#1}
%%%%%%%%%% Prefix a "S" to all equations, figures, tables and reset the counter %%%%%%%%%%
\begin{widetext}
\
 \vspace{3cm}
 
\centering

{\huge Supplemental Material} 

\vspace{1.5cm}

{\LARGE ``\papertitle''}\\
\vspace{1cm}

{\Large Tomokatsu Onaga, Fabio Caccioli and Teruyoshi Kobayashi}

\begin{figure*}[h]
    \centering
    \includegraphics[width=17cm]{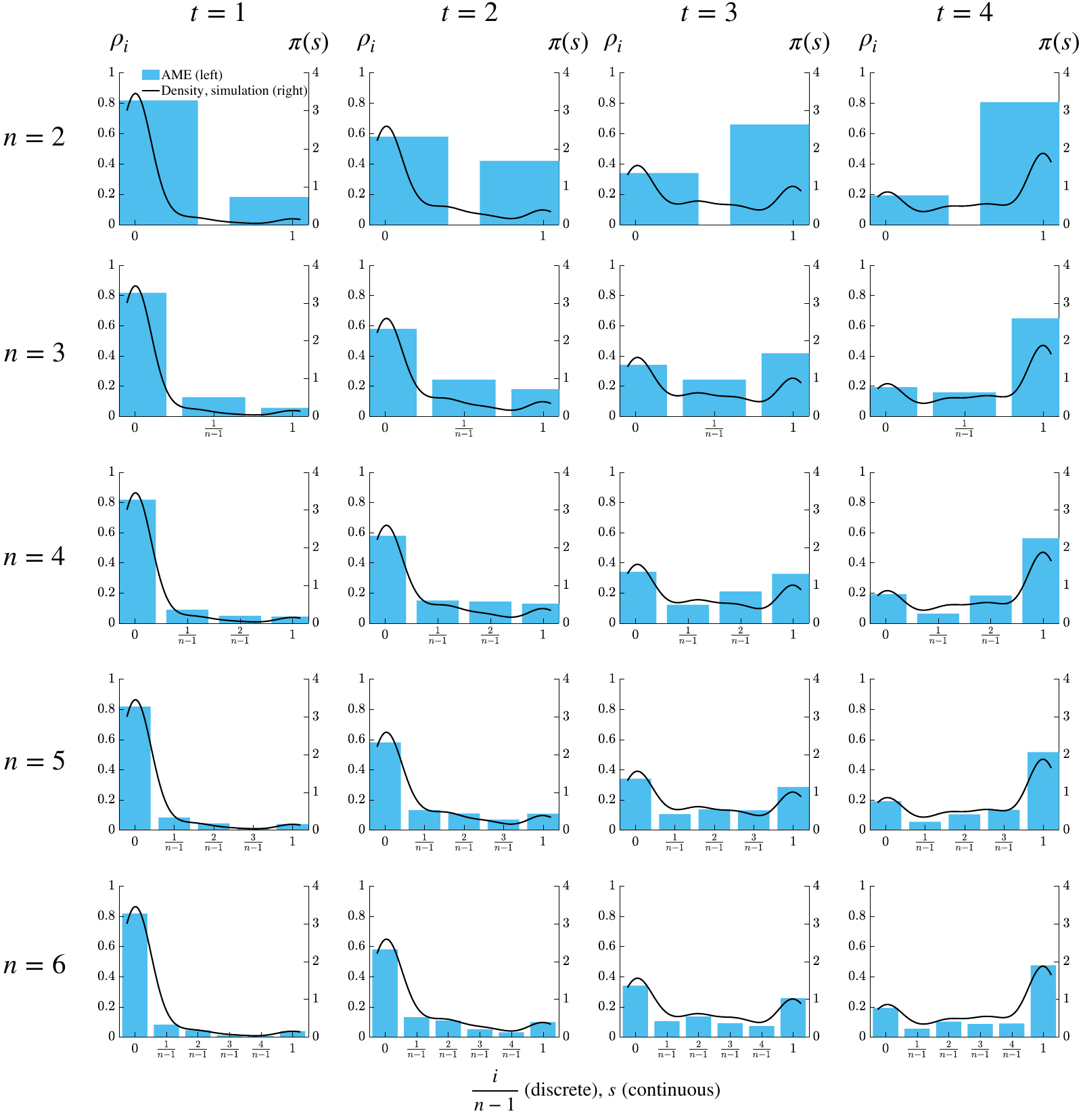}
    \caption{
    State distribution in the discrete- and continuous threshold models for different number of states $n$: response function~\eqref{eq:F_continuous_watts_nonlin}. In the discretized model, we use the AMEs to calculate the fraction of nodes in state~$i\in\{0,\ldots,n-1\}$, denoted by $\rho_i$ (blue bar). In the continuous-state model, the density $\pi(s)$ for state $s\in[0,1]$ is calculated by the kernel density estimation with bandwidth $0.1$ based on numerical simulations (black line). The index of discretized states is normalized by $n-1$ so that the states are distributed on $[0,1]$. Seed fraction is determined such that the fraction $0.05$ of nodes are in the ``most-active'' state (i.e., $i=n-1$ or $s=1$) while $0.95$ of nodes are in state~0. We run 40 simulations on \ER networks with $N=3000$, $z=3$, $\theta=0.18$, and $\eta=1$, where $\theta_j=\theta + (1-\theta)j/(n-1)$. Clearly, the larger the number of states, the better the matching between the simulation in the continuous-state model and its discrete approximation.}
    \label{fig:continuous_watts_theta018}
\end{figure*}

\begin{figure*}[h]
    \centering
    \includegraphics[width=17.5cm]{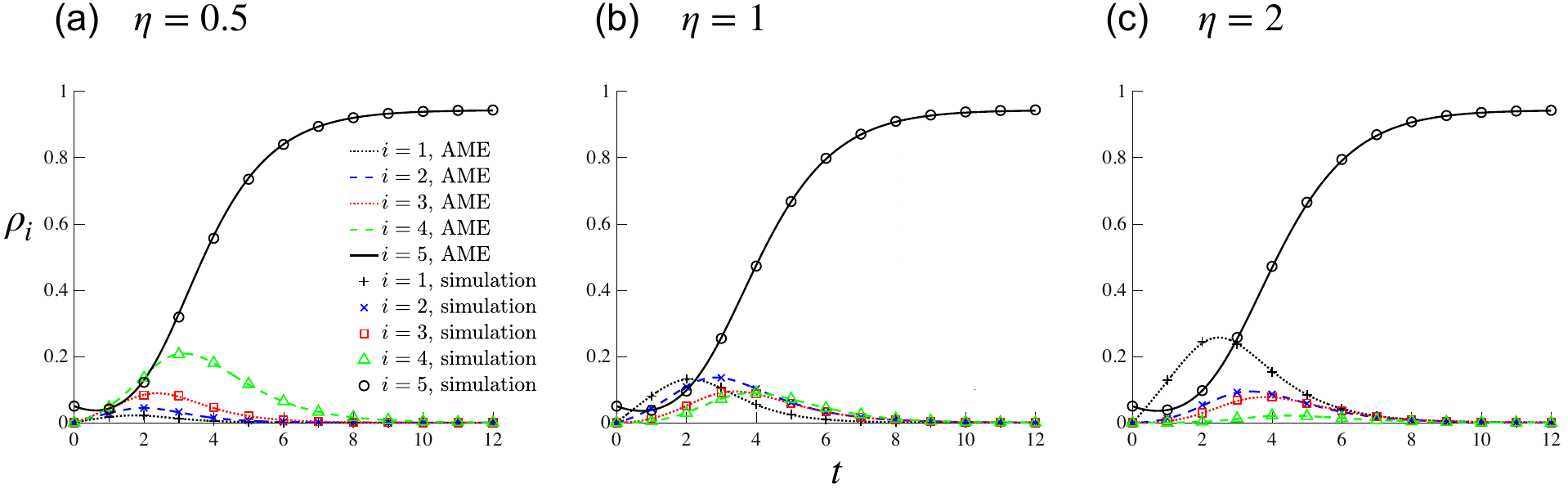}
    \caption{Dynamical path of the share of each state for different responsiveness $\eta$. The responsiveness is (a) progressive $(\eta<1)$, (b) neutral $(\eta=1)$, and (c) conservative $(\eta>1)$.
    We run 40 simulations on \ER networks with $N=3000$, $z=3$, $n=6$, $\theta=0.18$, where $\theta_j$ is given by Eq.~\eqref{eq:discretized_threshold_SI}. See the caption of Fig.~\ref{fig:continuous_watts_theta018} for the seed size.  }
    \label{fig:path_different_eta}
\end{figure*}

\begin{figure*}[h]
    \centering
    \includegraphics[width=17.5cm]{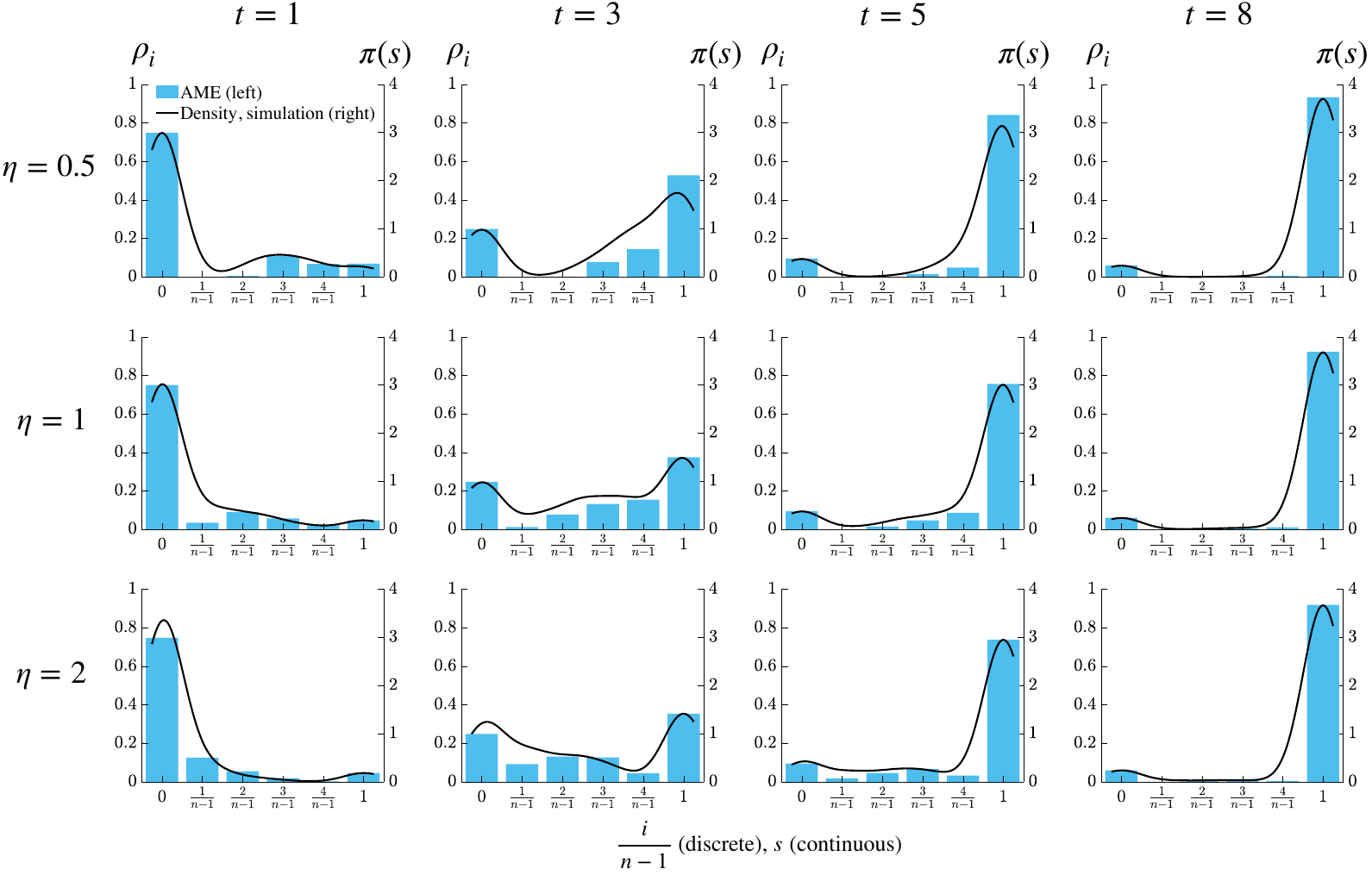}
    \caption{State distribution for different responsiveness $\eta$: response function~\eqref{eq:F_continuous_watts_nonlin}. We run 40 simulations on \ER networks with $N=3000$, $z=3$, $n=6$, $\theta=0$, where $\theta_j=\left(\frac{j}{n-1}\right)^\frac{1}{\eta}$. See the caption of Fig.~\ref{fig:continuous_watts_theta018} for the seed size. $\eta>1$ and $\eta<1$ respectively indicate conservative and progressive responses, while $\eta=1$ is neutral. }
    \label{fig:continuous_watts_dist_eta}
\end{figure*}

\begin{figure*}
    \centering
    \includegraphics[width=17cm]{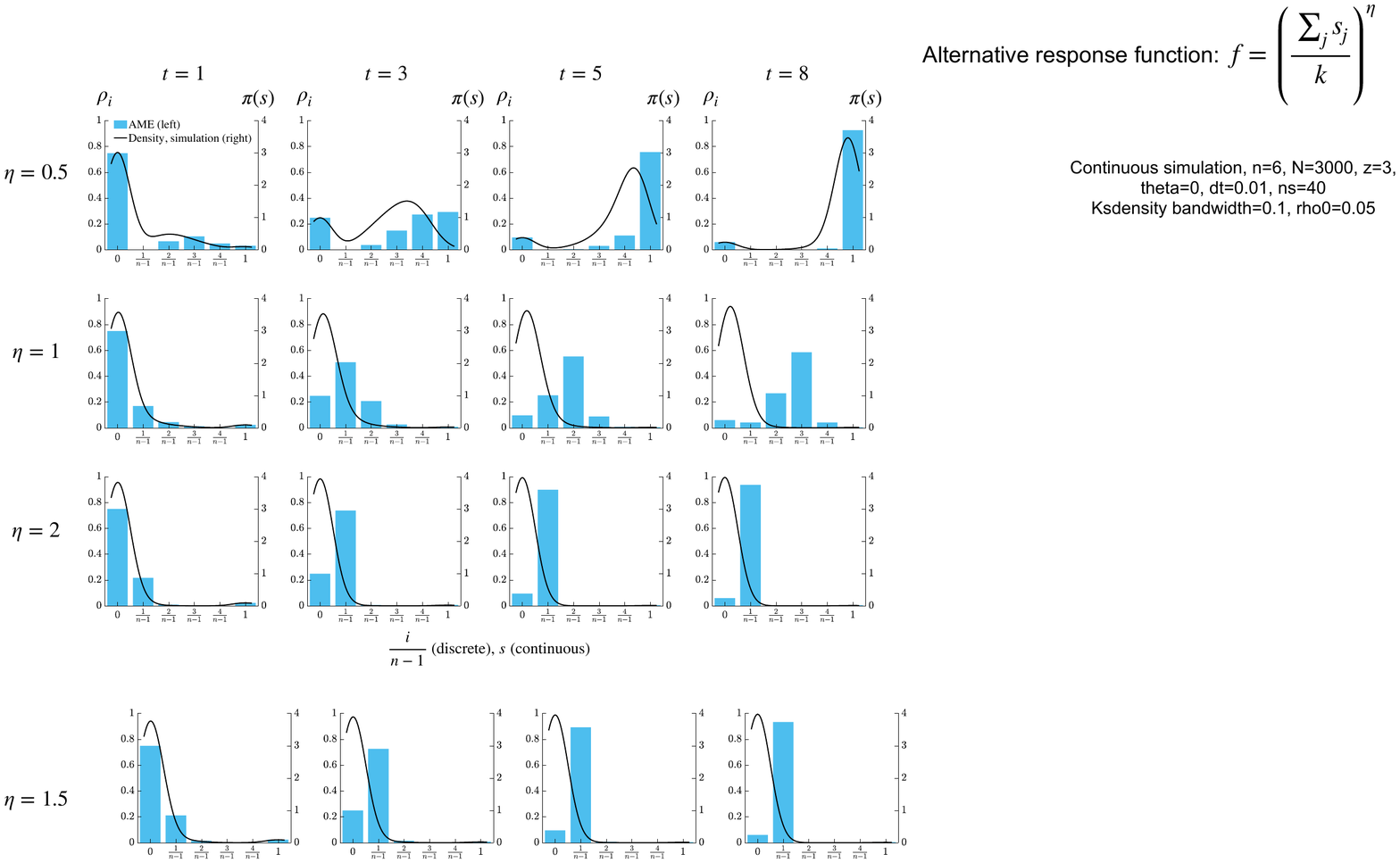}
    \caption{State distribution for different responsiveness $\eta$: response function~\eqref{eq:F_continuous_watts_nonlin_tilde}. We run 40 simulations on \ER networks with $N=3000$, $z=3$, $n=6$, $\theta=0$, where $\theta_j=\left(\frac{j}{n-1}\right)^\frac{1}{\eta}$. See the caption of Fig.~\ref{fig:continuous_watts_theta018} for the seed size.}
    \label{fig:continuous_AME_sum}
\end{figure*}

% \begin{figure*}[htb]
%     \includegraphics[width=15cm]{figs/AME_price_distribution.pdf}
%     \caption{Distribution of asset prices obtained by the AME ($n=6$) and simulation in the continuous-price model at a given point in time. See Fig.~\ref{fig:AME_comparison_benchmark} for the other parameter values.}
% \label{fig:AME_price_distribution} 
% \end{figure*}

\clearpage

\begin{figure}[htb]
\centering
    \includegraphics[width=15.8cm]{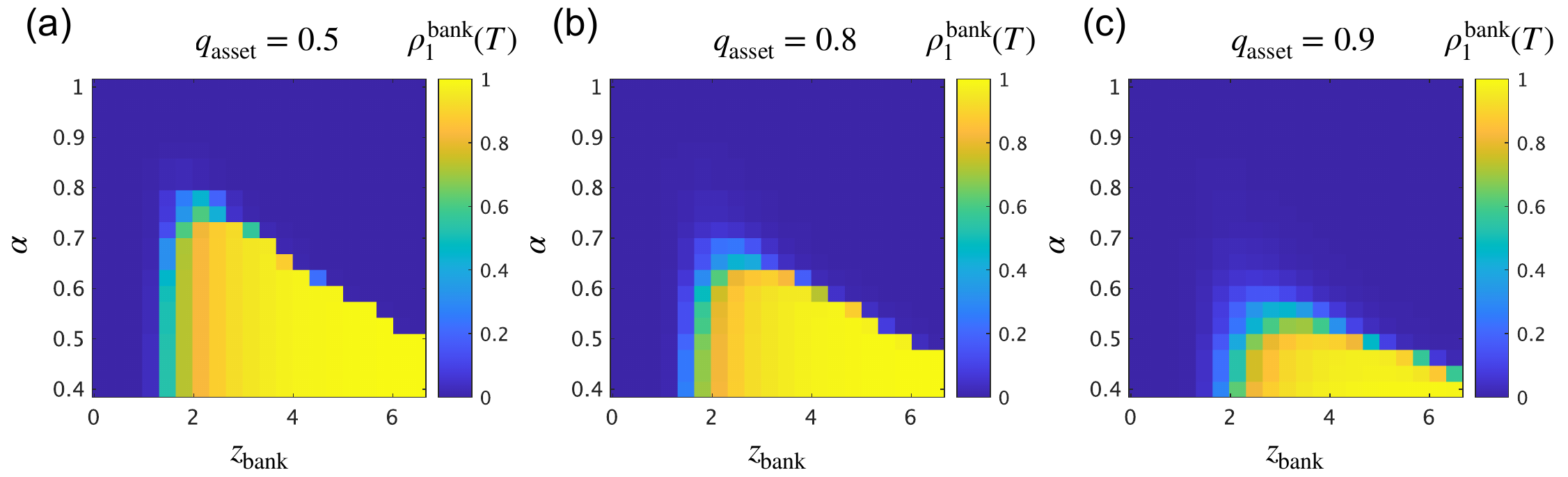}
    \caption{Cascade region under different degrees of policy intervention: asset-purchasing policy. We set $T=60$. See the caption of Fig.~\ref{fig:AME_comparison_benchmark} for the other parameter values.
    }
 \label{fig:ame_intervention_asset} 
\end{figure}
\bigskip

\begin{figure}[htb]
\centering
    \includegraphics[width=15.8cm]{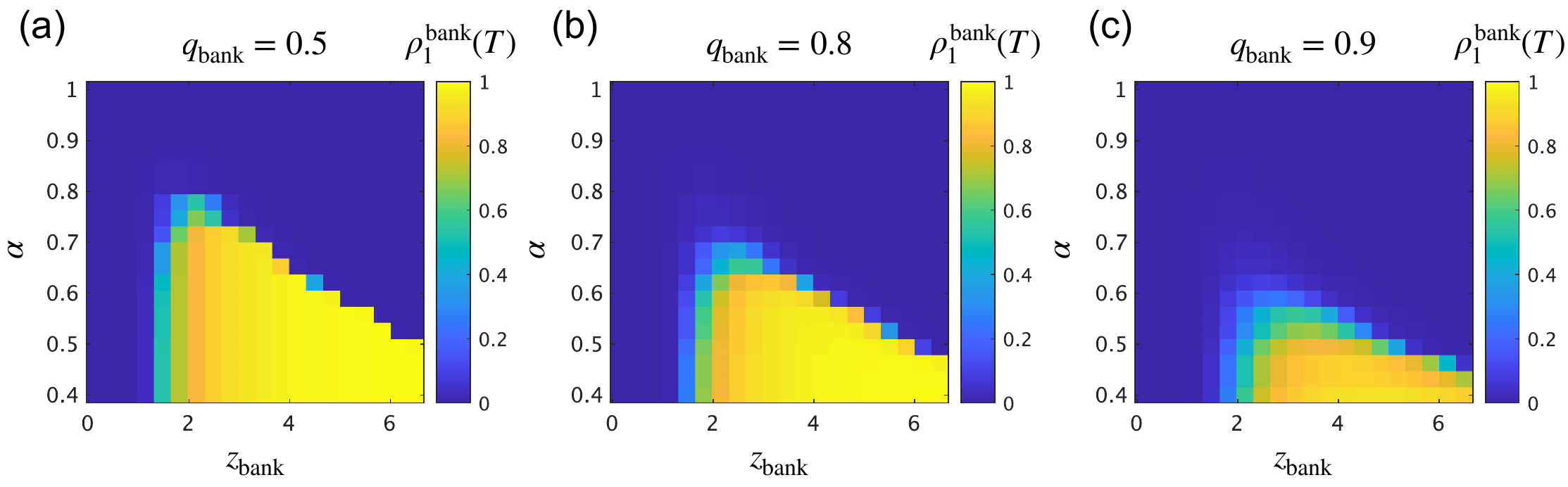}
    \caption{Cascade region under different degrees of policy intervention: bank-rescue policy. We set $T=60$. See the caption of Fig.~\ref{fig:AME_comparison_benchmark} for the other parameter values.}
    \label{fig:ame_intervention_bank}
\end{figure}

\end{widetext}

\end{document}